\documentclass[acmtog]{acmart}
\acmSubmissionID{1115}

\usepackage{booktabs} 

\usepackage{enumitem}

\usepackage[ruled]{algorithm2e} 

\SetAlFnt{\small}
\SetAlCapFnt{\small}
\SetAlCapNameFnt{\small}
\SetAlCapHSkip{0pt}
\usepackage[final,commandnameprefix=always]{changes}

\acmJournal{TOG}

\begin{document}
\title[NICER: A New and Improved Consumed Endurance and Recovery Metric]{NICER: A New and Improved Consumed Endurance and Recovery Metric to Quantify Muscle Fatigue of Mid-Air Interactions}


 \author{Yi Li}
 \affiliation{%
   \institution{Monash University}
   \city{Melbourne}
   \country{Australia}
 }
 \email{yi.li5@monash.edu}

  \author{Benjamin Tag}
 \affiliation{%
   \institution{Monash University}
   \city{Melbourne}
   \country{Australia}
 }
 \email{Benjamin.Tag@monash.edu}

  \author{Shaozhang Dai}
 \affiliation{%
   \institution{Monash University}
   \city{Melbourne}
   \country{Australia}
 }
\email{Shaozhang.Dai1@monash.edu}

 \author{Robert Crowther}
 \affiliation{%
   \institution{University of New England}
   \city{Armidale}
   \country{Australia}
 }
\email{rcrowth2@une.edu.au}

  \author{Tim Dwyer}
 \affiliation{%
   \institution{Monash University}
   \city{Melbourne}
   \country{Australia}
 }
\email{Tim.Dwyer@monash.edu}

  \author{Pourang Irani}
 \affiliation{%
   \institution{University of British Columbia}
   \city{British Columbia}
   \country{Canada}
 }
\email{pourang.irani@ubc.ca}

  \author{Barrett Ens}
 \affiliation{%
   \institution{University of British Columbia}
   \city{British Columbia}
   \country{Canada}
 }
 \email{barrett.ens@ubc.ca}
 \renewcommand{\shortauthors}{Li et al.}

\begin{abstract}
Natural gestures are crucial for mid-air interaction, but predicting and managing muscle fatigue is challenging. Existing torque-based models are limited in their ability to model above-shoulder interactions and to account for fatigue recovery. We introduce a new hybrid model, \textit{NICER}, which combines a torque-based approach with a new term derived from the empirical measurement of muscle contraction and a recovery factor to account for decreasing fatigue during rest. We evaluated NICER in a mid-air selection task using two interaction methods with different degrees of perceived fatigue. Results show that NICER can accurately model above-shoulder interactions as well as reflect fatigue recovery during rest periods. Moreover, both interaction methods show a stronger correlation with subjective fatigue measurement ($\rho = 0.978\slash 0.976$) than a previous model, Cumulative Fatigue ($\rho = 0.966\slash 0.923$), confirming that NICER is a powerful analytical tool to predict fatigue across a variety of gesture-based interactive applications.
\end{abstract}

%
%

%
%
\setcopyright{acmlicensed}
\acmJournal{TOG}
\acmYear{2024} \acmVolume{43} \acmNumber{4} \acmArticle{102} \acmMonth{7}\acmDOI{10.1145/3658230}


\maketitle

\section{Introduction}
Gesture Interaction has long been applied by Human-Computer Interaction (HCI) researchers to support direct manipulation~\cite{Elmqvist2011, Lee2021}, externalise cognition~\cite{NormanSmart}, facilitate remote collaboration~\cite{HarrisonVR}, and provide a ``natural'' user experience~\cite{BowmanNatural}. Multi-touch surface gestures have become the dominant mode of computer interaction and 3D mid-air gestures hold similar promise for the future of spatial computing~\chadded{\cite{lien2016soli,Taylor2016}}. Mid-air hand tracking is now supported by the newest generation of wearable Augmented Reality (AR) and Virtual Reality (VR) devices (e.g. Apple Vision Pro\footnote{\url{https://www.apple.com/apple-vision-pro/}}, Meta Quest 3\footnote{\url{https://www.meta.com/quest/quest-3/}}). Games that employ gesture input and fitness applications that use full-body gesture interaction are among the currently most popular VR experiences.

One vital aspect in designing such experiences is to present users with an engaging embodied experience without tiring them through excessive activity. Furthermore, the required level of exertion must be tuned to the specific application. For instance, to present a physical challenge when desired, but without inducing fatigue prematurely. Thus, it will be helpful for designers to have access to tools that can help them predict the exertion level of a particular activity and estimate the time the activity can be sustained.

Producing this type of information precisely is the goal of recent research on modelling fatigue in mid-air arm gestures. However, each of these approaches has particular benefits and drawbacks. For instance, the Consumed Endurance (CE) model~\cite{Hincapie-Ramos_CE_2014} was the first to predict endurance time but is limited in its ability to predict the effort of low-intensity activity that is typical of mid-air computer interactions. The Cumulative Fatigue (CF) model~\cite{jang2017modeling} was introduced soon after, with a proposed solution to the identified limitation in CE, as well as the ability to model fatigue recovery during interactions. However, the CF model takes a supervised learning approach to fit the model to subjective fatigue scales and, therefore, requires time-consuming model training on a specific interaction task and cannot directly predict the maximum interaction time.

Both of these previous models have a further inherent limitation in their assumption of shoulder torque as an accurate predictor of exertion in shoulder muscles which causes overestimation of fatigue in above-shoulder interactions. Recent work~\cite{revisitingCE2023} provides strong evidence countering this assumption, as well as a proposed solution, NICE, for addressing this limitation. This tentative model takes a \textit{hybrid approach} that supplements torque with direct observations of muscle activation, thereby adding a correction term for underestimated exertion for large shoulder angles. 

Our work \chreplaced{completes the previous exploration}{follows this previous research} using a `NICER' approach (\textbf{NICE} + \textbf{R}ecovery factor): We \chreplaced{propose a complete version of the theoretical correction term proposed in~\cite{revisitingCE2023} that improves fatigue prediction in over-shoulder exertion based on the empirical investigation of muscle fatigue development across different shoulder abduction angles in a long-duration endurance study and new arm movement data in unconstrained mid-air interactions.}{conduct a new endurance arm-lifting study with a design similar to the one in NICE~\cite{revisitingCE2023} but with a higher granularity of static shoulder poses (including seven angles between 45\textdegree\ and 135\textdegree\ versus only five angles in the literature). To that end, we establish the shape of the curve of objective muscle fatigue at different vertical shoulder positions. Next, a mid-air interaction task with constraint-free arm movement is run to obtain the precise coefficient of the curve and further refine the correction term initially proposed by NICE.} 

Additionally, we adapt a recovery factor that \chreplaced{was initially applied in industrial work to model fully relaxed breaks~\cite{ma2010new} in NICER to adjust for intermittent interactions for a more natural modelling of fatigue without a full stop of tasks. Ours is the first comprehensive model to }{allows us to present a complete, refined model that}address the full set of prior limitations (summarised in Table~\ref{tab:modelsummary} and explained in detail in Section~\ref{sec:NICER}). Lastly, we evaluate the model performance of NICER as an analytical tool for interaction design in another mid-air selection task with high- and low-fatigue interaction methods. The study results confirm that NICER is better able to differentiate between two interaction methods with different degrees of perceived fatigue as well as better capture of fatigue recovery than the previous models, CE and CF.

The main contribution of this work is a refined fatigue model that is ready-to-use\footnote{The model source code is available at \url{https://github.com/ylii0411/NICER_Unity_API}} (does not require any further model training), applicable to any arm pose (from hand, wrist, elbow, and shoulder joint coordinates), and comprehensive (addresses the limitations of the original CE and CF models outlined in Section~\ref{sec:NICER}). Specific contributions include:

\begin{itemize}

\item An empirical study of muscle fatigue development: the first investigation at different vertical arm positions with low-intensity, long-duration tasks (Study 1 in Section \ref{sec:study1}). 

\item A finalised comprehensive fatigue model: NICER that accounts for additional exertion of above-shoulder exertion and recovery periods (Study 2 in Section \ref{sec:model_development}).

\item An evaluation study to confirm that NICER is a reliable objective fatigue measurement to assist future interaction design (Study 3 in Section \ref{sec:model_evaluation}).
\end{itemize}

\section{Related Work}
Fatigue is a cumulative phenomenon that manifests after prolonged sustained physical exertion. In the following section, we discuss the literature on both subjective and objective fatigue assessments and the development of existing fatigue models in mid-air interactions. It is necessary to clarify that this paper focuses exclusively on physical fatigue, specifically muscular fatigue, with mental fatigue being outside the scope of our study.

\subsection{Subjective Fatigue Measurement}
Subjective approaches to quantify perceived fatigue are common in Human-Computer Interaction (HCI) as they require no physical instrumentation or additional setup. Two widely accepted subjective techniques are the Borg CR10 Rating of Perceived Exertion (RPE) scale~\cite{borg1998borg} and the National Aeronautics and Space Administration Task Load Index (NASA-TLX)~\cite{hart1988development}. While the Borg CR10 RPE employs discrete values ranging from 0 to 10 to categorise physical exertion, the NASA-TLX evaluates perceived workload by considering dimensions such as physical demand, mental demand, and frustration using a 21-point scale. Borg CR10 RPE and NASA-TLX provide a relatively reliable rough estimation of fatigue levels. However, two noteworthy issues limit their application in managing fatigue during mid-air interactions. First, the finite scales in the subjective approach restrict their performance in detecting subtle yet crucial differences~\cite{Hincapie-Ramos_CE_2014}. Second, participants with diverse backgrounds may interpret and apply the scales differently and introduce potential bias to the results~\cite{borg1998borg, Kosch2023}. Furthermore, completing the questionnaire is disruptive to the activity under investigation and is not practical in applied settings.

\subsection{Objective Fatigue Measurement}
\label{sec:objective_fatigue}
Compared to subjective fatigue approaches, objective measurements involve monitoring physiological properties, e.g., heart rate~\cite{barclay2011quantitative}, muscle oxygenation~\cite{ferguson2011shoulder}, thermoregulation~\cite{Aryal2017}, and eye-movement~\cite{Tag2019}. 

While objective measurements eliminate the influence of cognitive bias across participants, they are inferential rather than direct indicators of muscle fatigue and could introduce noise into the fatigue assessment process. For example, lightweight wearable devices like smartwatches or fitness trackers allow real-time heart rate monitoring. Yet, the subtle variation in heart rate during low-intensity tasks may be caused by many factors other than fatigue induced by interaction~\cite{barclay2011quantitative} (e.g., it can even be induced by the subject's awareness of displayed heart rate, i.e.\ the well-known \textit{observer effect}).

A non-invasive option to estimate participants' fatigue is through their expended muscle capacity~\cite{hayes2002reliability}. Force transducers record the Maximum Voluntary Contraction (MVC) value representing the maximum muscle contraction capacity for a given participant. Since muscle capacity is reduced over the course of a physical task, a decline in MVC over time serves as an indicator of muscle fatigue. However, this direct approach requires additional setup (i.e.\ measuring muscle capacity with a force transducer) that will interrupt the interaction, so it fails to provide feedback on fatigue during real-time interactions. 

Similarly, MVC can be measured directly through surface electromyographic (EMG) signals. After normalization by MVC, the real-time sub-maximal EMG signals during interactions will be expressed in the unit of \%MVC. An increasing magnitude in time-domain features of EMG signals (i.e. root-mean-square (RMS), mean-absolute-value (MAV)) is considered an indicator of muscle fatigue~\cite{yousif2019assessment}. Consequently, slopes of the time-domain features (\%MVC) are applicable as fatigue indices to monitor changes in muscle contraction in real-time interactions~\cite{naik2014applications}. This approach is recommended if the study duration is long enough to observe a longitudinal linear change in EMG signal above the level of intermittent noise. \chadded{Training classification models on instantaneous EMG signals is also feasible to predict fatigue~\cite{wang2021muscle}, but this approach is limited to discrete fatigue levels and may overlook the subtle difference between scales.}

\subsection{Modelling Fatigue in Mid-air Interaction}
In the field of HCI, modelling fatigue in mid-air interactions is an emerging area, building upon established research in ergonomics. \cite{plantard2015pose} integrated the widely-recognized Rapid Upper Limb Assessment (RULA) ergonomic metric with a markerless motion tracking system, facilitating real-time assessment of ergonomic costs. \cite{bachynskyi2014motion, bachynskyi2015informing} sectioned the 3D interaction volume, utilizing biomechanical simulations to estimate muscle activation costs for various interaction clusters. More recently, an AR toolkit named Xrgonomics~\cite{Belo_XRgonomics_2021} incorporated these metrics to inform 3D user interface design. While these prior works contribute to reducing muscle fatigue by minimizing ergonomic costs, their focus has primarily been on evaluating static postures, rather than dynamic movements that involve changes in gestures, velocity, and acceleration.

\begin{figure}[h]
    \centering
    \includegraphics[width=0.48\textwidth]{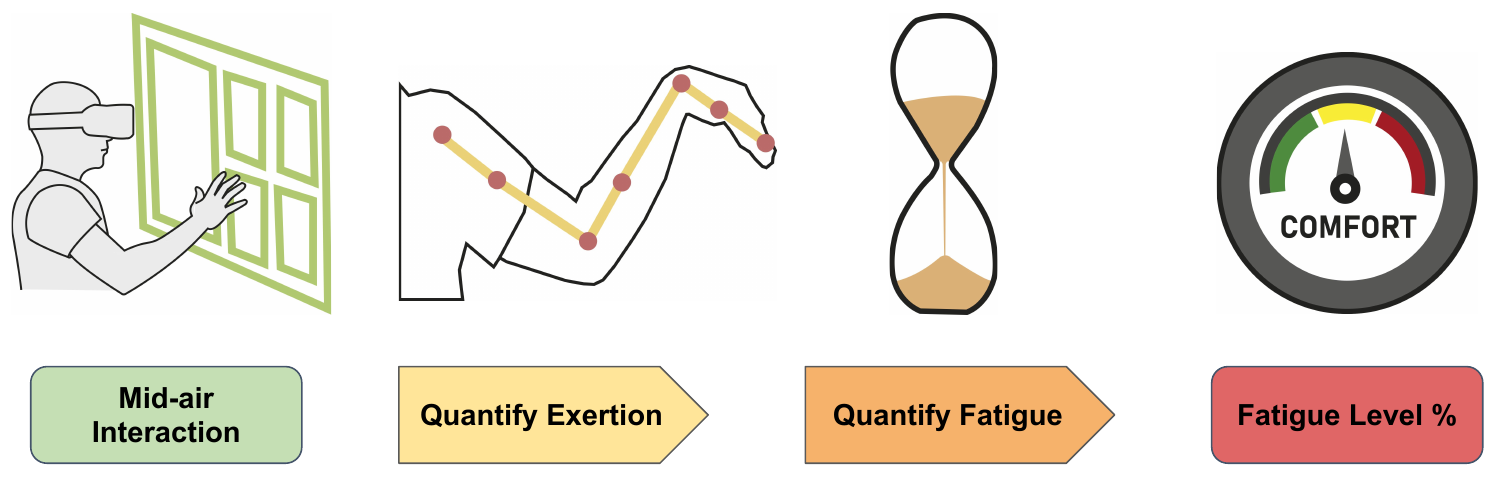}
\caption{Fatigue modelling pipeline for mid-air interaction, requires differentiating two key measures: \textit{exertion} and \textit{fatigue}.}
    \label{fig:FatiguePipeline}
\end{figure}

The process of modelling physical fatigue in mid-air interaction requires defining two measures (Fig~\ref{fig:FatiguePipeline}):

\noindent\textbf{Exertion} -- measuring instantaneous physical exertion of body gestures performed during the interaction (\%MVC); and 

\noindent\textbf{Fatigue} -- a mapping from the accumulated physical exertion (\%MVC) to a fatigue level (\%).

In general, there are empirical and theoretical approaches to model muscle fatigue. Empirical models rely on the observation of maximum holding time, A.K.A. endurance time (ET), or MVC. They consider the decay of ET or MVC to be equivalent to fatigue development. \cite{rodriguez2002joint} defined a lightweight fatigue model that can estimate fatigue levels by normalising actual holding time with the maximum holding time of a specific joint as the denominator. Even though the Rodriguez model has reduced parameters and considers resting periods compared to previous models, it cannot be applied to body segments containing multiple joints. Another empirical study by \cite{ma2009new,ma2010new} effectively estimated ET by modelling the decay coefficient of muscle capacity for a given external load. However, the model is limited to isometric tasks, a type of continuous contraction that does not require changes in body postures. 

On the other hand, theoretical fatigue models quantify muscle fatigue based on the biophysical properties of the muscle fibres. The dynamic fatigue model proposed by \cite{liu2002dynamical} was the first study to model fatigue by considering the transition cycle of the muscle unit, where the proportion of the fatigued muscle units (between 0-100\%) can be estimated for a constant intensity. The Liu model was the predecessor of the three-compartment model (TCM)~\cite{xia2008theoretical,frey2012three,frey2021muscle, looft2020adapting} that enables the muscle transition cycle to work with varying external loads.

\begin{figure}[h]
    \centering
    \includegraphics[width = \linewidth]{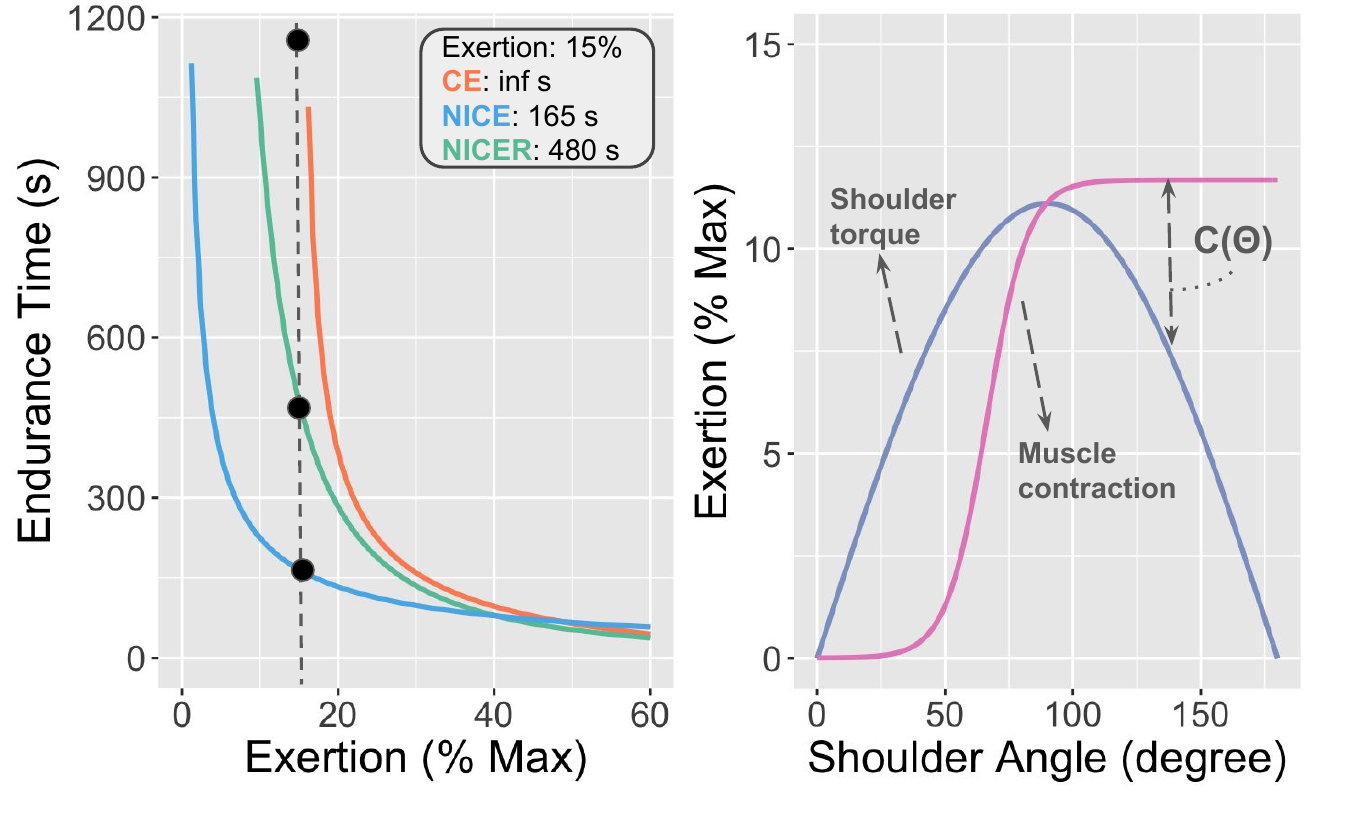}
    \caption{Left: The ET function implemented in CE (orange), NICE (blue), and NICER (green). Right: The instantaneous exertion measured in shoulder torque (purple) and muscle contraction (pink) under different shoulder angles. }
    \label{fig:EMG_Torque}
\end{figure}

The above literature laid a solid foundation for the real-time fatigue modelling process and inspired the development of two milestone fatigue models of mid-air interactions: Consumed Endurance (CE)~\cite{Hincapie-Ramos_CE_2014} and Cumulative Fatigue (CF)~\cite{jang2017modeling,villanueva2023advanced}\footnote{CF, as the model name, refer to the "TCM" model in~\cite{jang2017modeling} and the "LIN" model in~\cite{villanueva2023advanced}.}. CE was the first model that successfully estimated fatigue from dynamic arm movements as described in the pipeline (Figure\ref{fig:FatiguePipeline}). It first quantifies the instantaneous exertion at time $i$ by the shoulder torque ($Torque_{i}$) based on the performed gesture, then normalises the accumulative average torque ($Torque$) with a constant $Max\_Torque$ retrieved from literature. Rohmert's ET function (Equation (\ref{eq:rohmertET})) is applied next to map the accumulative exertion to its maximum holding time (see the orange curve in Figure~\ref{fig:EMG_Torque}). The $CE$ score is predicted by the ratio between the $SpentTime$ and the estimated $ET_{Rohmert}$ from Rohmert's ET function (see Equation (\ref{eq:CE_final})). 

\begin{equation}
\label{eq:rohmertET}
    ET_{Rohmert} = \frac{1236.5}{(\frac{Torque}{Max\_Torque} *100 - 15 )^{0.618}} -72.5
\end{equation}

\begin{equation}
\label{eq:CE_final}
    CE = \frac{SpentTime}{ET_{Rohmert}} *100
\end{equation}

In an initial evaluation, CE was strongly correlated with Borg CR10 and demonstrated the ability to guide interaction designs \cite{Hincapie-Ramos_CE_2014}. However, CE has a major limitation when working with low-intensity interactions due to the biased assumption of Rohmert's ET function \chadded{(the orange curve in Figure~\ref{fig:EMG_Torque}-left)} that low-intensity interaction \chadded{(i.e. less than 15\% of an individual's maximum strength)} is considered to last forever without subjects becoming fatigued. This was addressed in NICE~\cite{revisitingCE2023} where Rohmert's ET function was replaced by a revised ET function constructed from empirical evidence \chadded{(the blue curve in Figure~\ref{fig:EMG_Torque}-left)}. \chadded{However, the revised ET function in NICE only considered static shoulder positions and overlooked dynamic arm gestures. In our proposed improved model, we overcome the previous limitation by using a shoulder-specific ET function (the green curve in Figure~\ref{fig:EMG_Torque}-left) built from a meta ET function fitted to both static and dynamic arm gesture data from literature~\cite{frey2010endurance}.} 

Alternatively, \chadded{the original CE uses a constant $Max\_Torque$ $(N\cdot m)$ to normalise instantaneous shoulder torque $Torque$ $(N\cdot m)$ to $Torque$ $(\%MVC)$. Yet, this lightweight approach ignores variations between participants.} The original CF~\cite{jang2017modeling} used a long-duration arm-lifting task to determine one's maximum strength (MVC), but this pre-calibration was time-consuming and may induce fatigue before the interaction. The pre-calibration was recently replaced by a gesture-based maximum strength model~\cite{chaffin2006occupational} implemented in the revised CF~\cite{villanueva2023advanced}. \chadded{The same solution is included in our proposed improved model to capture variations between participants and dynamic gestures. See APPENDIX~\ref{sec:chaffinmodel} for more details.}

Furthermore, CE and NICE do not consider rest periods during interactions. As such, model predictions will overestimate the fatigue level when an interaction is resumed after a break. Concerns related to fatigue recovery in interaction were explored in CF~\cite{jang2017modeling, villanueva2023advanced} where a TCM was used to estimate the proportion of fatigued ($F\%$) and resting ($R\%$) muscle units under the current shoulder torque of given gestures. Relying on a supervised-learning approach where CF model parameters (fatiguing and recovery rates) were trained to fit Borg CR10 collected from the interaction study, CF predicts subjective fatigue ratings with low error. However, predicting Borg CR10 limits CF in estimating the ET of interaction. 

Most importantly, CE and CF quantify the instantaneous exertion solely based on shoulder torque, which assumes the exertion below and above the shoulder is of the same level. Yet, this assumption is contradicted by subjective feedback~\cite{Belo_XRgonomics_2021} and measured muscle contraction \cite{revisitingCE2023}. In NICE, an additional logarithmic term was added to the original torque calculation and an improved correlation with muscle contraction in above-shoulder interactions was observed. Though the approach was not validated, the work pointed out that a hybrid approach combining torque and muscle contraction may be able to effectively quantify the additional exertion in above-shoulder interactions. This realisation, along with the above concerns, motivated our creation of a comprehensive fatigue model that can:
\begin{enumerate}
    \item \textbf{work with interactions of varying physical intensities}, 
    \item \textbf{consider gesture-based maximum strength},
    \item \textbf{account for fatigue recovery during interaction breaks},
    \item \textbf{predict the endurance time of interactions}, and
    \item \textbf{make reliable predictions for above-shoulder interactions}.
\end{enumerate}

\section{NICER: Nice and Improved Consumed Endurance and Recovery Metric}
\label{sec:NICER}
Our work aims to develop a ready-to-use comprehensive fatigue model for mid-air interactions. We investigated the strengths and limitations of the existing fatigue models in the literature and summarised them in Table~\ref{tab:modelsummary}. We plan to integrate four new features into the original CE model to develop a comprehensive model that preserves the advantages of existing models and overcomes the identified limitations related to above-shoulder exertion. The order of the proposed changes is based on the computation order of the original CE formula: (1) A revised ET curve to enable the refined model to work with low-intensity interactions; (2) A correction term derived from muscle contraction to account for the additional physical effort needed in above-shoulder interactions (Section~\ref{sec:hybridmodel}); (3) Chaffin's maximum strength model to replace the constant maximum torque to account for the variation in the maximum strength of dynamic gestures;
~(4) A recovery factor to reflect decreasing fatigue during rest periods (Section~\ref{sec:recoveryfactor}).

Next, we discuss the detailed motivation and methodology for \chreplaced{our novel hybrid exertion estimation and the recovery factor}{each new model feature}, as well as the model evaluation in the following sections.


\begin{table*}[]
\caption{Comparison of existing fatigue models against our proposed NICER model, listing features supported by each. Y: feature supported, N: not supported.}
\begin{tabular}{|c|c|c|c|c|c|c|}
\hline
\begin{tabular}[c]{@{}c@{}}Fatigue\\ Model\end{tabular} & \begin{tabular}[c]{@{}c@{}}Work with \\ low-intensity \\ interactions\end{tabular} & \begin{tabular}[c]{@{}c@{}}Apply \\ gesture-based \\ maximum \\ strength\end{tabular} & \begin{tabular}[c]{@{}c@{}}Consider \\ fatigue \\ recovery\end{tabular} & \begin{tabular}[c]{@{}c@{}}Predict \\ endurance \\ time\end{tabular} & \begin{tabular}[c]{@{}c@{}}Account for \\ above-shoulder \\ exertion\end{tabular} & \begin{tabular}[c]{@{}c@{}}Evaluation in \\ interaction tasks\end{tabular} \\ \hline
CE~\cite{Hincapie-Ramos_CE_2014} & N & N & N & Y & N & Y \\ \hline
CF\_2017~\cite{jang2017modeling} & Y & N & Y & N & N & Y \\ \hline
CF\_2023~\cite{villanueva2023advanced} &Y & Y & Y & N & N & Y \\ \hline
NICE~\cite{revisitingCE2023} & Y & N & N & Y & Y & N \\ \hline
NICER & Y & Y & Y & Y & Y & Y \\ \hline
\end{tabular}
\label{tab:modelsummary}
\end{table*}



\subsection{Hybrid Approach of Real-time Exertion Estimation}
\label{sec:hybridmodel}

CE and CF rely on shoulder torque to estimate the instantaneous exertion of the currently performed arm gesture. However, the property of shoulder torque\footnote{$Torque = r*F*sin(\theta)$. $\theta$ is the angle between the upper arm and the body, $r$ is the distance between the shoulder joint and the centre of the mass of the arm, and $F$ is the force exerted by the arm.} 
suggests that exertion is symmetrical at equal shoulder angles above and below 90\textdegree{}  (see the purple curve in Figure~\ref{fig:EMG_Torque}-right). 

In the evaluation of CE and CF in mid-air pointing tasks, subjective feedback of perceived fatigue~\cite{liu2018applying,Belo_XRgonomics_2021,revisitingCE2023} and objective muscle contraction through EMG signals~\cite{revisitingCE2023} indicated that CE and CF overlooked the additional physical effort needed for lifting the arm above 90\textdegree\ and therefore underestimated the fatigue level induced by above-shoulder interactions.

NICE was the first study investigating the limitations of only using torque to quantify instantaneous exertion of shoulder movements, and it proposed a correction term for exertion estimation based on torque measurements at 120\textdegree\ and 90\textdegree\ of shoulder elevation, assuming a linear increase in exertion between these elevations. However, this assumption has not been tested previously.

On the contrary, using solely muscle contraction to predict shoulder fatigue requires extensive mappings between muscle activation and properties of dynamic interactions, for example, contributing shoulder muscles, arm movement velocity, interactive target positions, and so on. Unfortunately, research done in the above areas is limited.

The refined fatigue model proposed in this paper takes a hybrid approach, integrating muscle contraction and torque to address the identified gap. We use the mapping between shoulder angles and muscle fatigue as the objective ground truth to correct the torque calculation. The development of the correction term takes three steps:

\begin{enumerate}
    \item Obtain the shape of the curve of objective muscle fatigue under varying shoulder angles from low-intensity, long-duration tasks (i.e.\ 10 minutes per task);
    \item Investigate the magnitude of shoulder torque under continuously varying shoulder angles in constraint-free mid-air arm movement;
    \item Develop the correction term based on the difference between shoulder torque and the ground truth curve.
\end{enumerate}

As introduced in Section~\ref{sec:objective_fatigue}, EMG signals with an increasing magnitude for a given period indicate the occurrence of muscle fatigue. 
The slope of EMG features reflects the severity of fatigue, such that a steeper slope indicates more rapid fatigue development. In Study 1 (Section~\ref{sec:study1}), we investigate objective fatigue from the distribution of EMG-based fatigue indices at different vertical shoulder angles. 

Later, in Study 2 (Section~\ref{sec:model_development}), we use data collected during mid-air interaction tasks with no constraints on shoulder abductions and elbow extensions to obtain the distribution of shoulder torque under continuously varying shoulder angles.

Finally, we scale up the ground truth curve to match the magnitude of shoulder torque and take the difference between the ground truth curve and the shoulder torque distribution to obtain the correction term $C(\theta)$ in Figure~\ref{fig:EMG_Torque}-right.

\chdeleted{Chaffin’s Model of Maximum Strength Estimation}

\chdeleted{In the original CE, a constant $Max\_Torque$ $(N\cdot m)$ was used to normalise instantaneous shoulder torque $Torque$ $(N\cdot m)$ to $Torque$ $(\%MVC)$. Yet, this lightweight approach ignored the variation between participants. Conversely, the original CF estimated participants’ $Max\_Torque$ through a long-duration arm lifting task, and the results showed a strong correlation with the direct measurement. Regardless, this pre-calibration process is time-consuming and can cause muscle fatigue before the interaction tasks. A potential solution is to estimate the $Max\_Torque$ based on kinematic information like velocity and joint angles. However, prior works have so far only explored such techniques in the knee~\cite{frey2012knee}, elbow~\cite{frey2012knee}, and wrist~\cite{xia2015wrist}, but not in the shoulder. An alternative approach is Chaffin's strength model~\cite{chaffin2006occupational} implemented in the revised CF~\cite{villanueva2023advanced}. Chaffin's model outputs the current $Max\_Torque$ $(N\cdot m)$ at the shoulder joint based on the performed arm gestures represented in the elbow extension angle ($\alpha_{e}$) and shoulder abduction angle ($\alpha_{s}$) while considering the difference of physical capacity between gender by factor $G~(G_{female} = 0.1495, G_{male} = 0.2845)$. The full model formula is shown in Equation~\ref{eq:chaffinmodel}.}


\chdeleted{In a direct comparison of the revised CF~\cite{villanueva2023advanced} with the original  CF model~\cite{jang2017modeling}, adding Chaffin's model significantly reduced the error between CF predictions and Borg CR10. Therefore, we include Chaffin’s model in NICER to capture variations between participants and dynamic gestures.}

\subsection{Recovery Factor to Account for Rest Periods}
\label{sec:recoveryfactor}

During a continuous mid-air interaction, the original CE outputs increasing model predictions to reflect the growing cumulative fatigue. However, the model prediction will become constant when the interaction is paused for a short break. 

Meanwhile, CF follows the life cycle of muscle units and considers muscles transitioning between fatigued and resting states during the entire periodic interaction. This conservative approach leads to an inconsistent performance of CF between active interaction periods and rest periods. In an initial validation~\cite{jang2017modeling}, CF achieved a smaller error with Borg CR10 during the interaction break and reported a higher error during active target-pointing tasks. A similar observation was found in applying CF in a multi-touch display, where the error between CF predictions and Borg CR10 is significantly higher during the active interaction than during the study break~\cite{liu2018applying}. The above concern highlights the need to consider fatigue recovery separately for active and rest periods.

A revised recovery factor was proposed by \cite{looft2018modification} where an additional scalar was applied to increase the recovery factor when the current exertion is low. However, this approach did not reduce the error between the fatigue prediction and the subjective ground truth and has only been evaluated in simulation data~\cite{cheema2020predicting,cheema2023discovering}.

In an early exploration of modelling shoulder and elbow fatigue for static physical tasks, \cite{ma2010new} evaluated the effectiveness of applying a recovery factor $R = 0.04~(s^{-1})$ to estimate the extended endurance after 30s breaks. The same recovery factor and the transition function (see Equation (\ref{eq:recoveryfactor})) from this prior work~\cite{ma2010new} are used in NICER to reduce the estimated fatigue level ($Fatigue_{t}$) at time $t$ when the shift between active interaction and the rest period is detected.

\begin{equation}
\label{eq:recoveryfactor}
    Fatigue_{t} (\%) = Fatigue_{t-1} (\%) \cdot \exp ^{-R \cdot \delta t}
\end{equation}

\subsection{Evaluating NICER as An Interaction Analytical Tool}
\label{sec:proposed studies}

We evaluate the refined model NICER against CF as interaction analytic tools by comparing them with subjective interaction measures in a mid-air selection task with different degrees of perceived fatigue. Three evaluation criteria are considered:
\begin{enumerate}
    \item whether fatigue models can generalise to different interaction designs, 
    \item whether fatigue models can reflect fatigue recovery during breaks in activity, and
    \item whether fatigue models can accurately predict fatigue levels.
\end{enumerate}

This evaluation strategy was inspired by the three evaluations done in CE~\cite{Hincapie-Ramos_CE_2014}. CE was evaluated under different mid-air interaction factors, and the model predictions showed agreement with Borg CR10 and completion time.
Meanwhile, CF~\cite{jang2017modeling, villanueva2023advanced} has yet to be evaluated as such a tool by being compared to objective measures rather than subjective Borg CR10 scores. CF may perform better than NICER if we evaluate the model performance based on the error between fatigue predictions and Borg CR10 scores. Doing so would require a conversion from NICER predictions to Borg CR10 scores. The original CF evaluation~\cite{jang2017modeling} proposed a linear factor for similarly converting CE to Borg CR10, however, they noted that such an approach has not been validated.

On the other hand, prior studies~\cite{Hincapie-Ramos_CE_2014, jang2017modeling, villanueva2023advanced} have confirmed that both CE and CF can differentiate interaction designs based on different target placements below the shoulder level. This differentiation relies on a torque-based approach to quantify exertion. Subsequently, the limitations of a torque-based approach were empirically confirmed in a study with above-shoulder target placements~\cite{revisitingCE2023}. Our work seeks to further explore and assess the performance of NICER and CF in studies of varying durations. CE predicts fatigue levels by estimating the maximum interaction duration, whereas CF bases its fatigue predictions on estimating the proportion of fatigued muscle units. 

Given the above concerns, we evaluate our proposed comprehensive model (derived from the results of Studies 1 and 2) against the three criteria listed above. Our evaluation presented in Study 3 (Section~\ref{sec:model_evaluation}) uses data collected from an unconstrained mid-air interaction task similar to Study 2. In this case, we contrast two interaction methods known to create different levels of subjective fatigue in a task with controlled rest periods. In addition to testing each model's prediction of fatigue recovery, this allows us to test their capability in differentiating between tasks requiring higher or lower levels of relative exertion, thus demonstrating the model's ability to generalise to unknown conditions.

\begin{figure}[h]
    \centering
    \includegraphics[width = 1\linewidth]{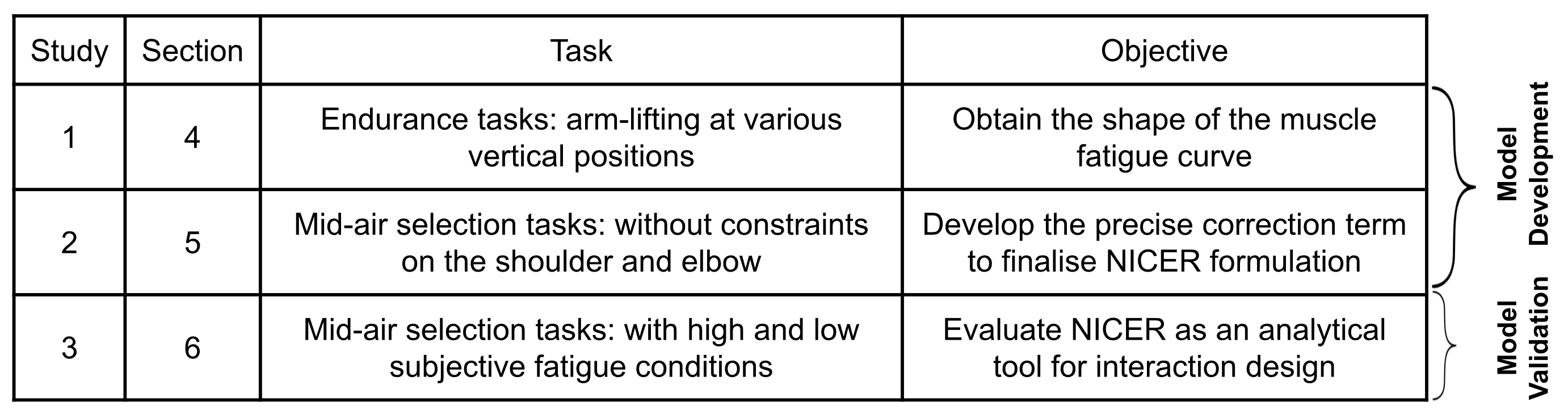}
    \caption{A summary of three studies conducted in the current paper.}
    \label{fig:study_summary}
\end{figure}

A summary of all three studies conducted in the current paper can be seen in Figure~\ref{fig:study_summary}.

\subsection{Dataset Used for Analyses in Studies 2 and 3}
\label{sec:StudyDataSet}
Whereas previous studies~\cite{Hincapie-Ramos_CE_2014, jang2017modeling, villanueva2023advanced} have implemented abstract tasks (e.g. text entry, docking) to test the ability of the proposed models to generalise, we aimed to evaluate our model with a more complex, externally valid task, with minimal constraints on the user's arm movement. We thus identified a task from a separate research study~\cite{magicportal2023} on interaction methods for mid-air selection of targets from a 3D scatterplot visualisation as it meets several criteria for our evaluation: 
\begin{enumerate}
    \item The controlled study is run in an immersive environment using a commodity VR display.
    \item The target selection task allows unconstrained shoulder motion to select targets in a large 3D volume.
    \item The task is unconstrained in time, providing interactions of varying time duration.
    \item The study was designed to measure task fatigue using subjective measures.
    \item The study involved multiple interaction methods, two of which were shown to cause statistically significant differences in subjective fatigue.
\end{enumerate}

Our analyses in Studies 2 and 3 use the same target selection tasks with multiple different interaction methods (four in total). Data from two of these interaction methods were used for refinement of our model in Section~\ref{sec:model_development}. The two remaining methods, which were shown to induce different levels of subjective fatigue (high and low) were used for model evaluation in Section~\ref{sec:model_evaluation}. The study descriptions in Studies 2 and 3 include only relevant details. A summary of the task and interaction methods is in Appendix~\ref{sec:AppendixDataSet}. Although we use the complete dataset, the assigned variables in our study design differ from those of the original study.

\section{Study 1: Understanding muscle fatigue at different shoulder abduction angles}
\label{sec:study1}

As explained in Section~\ref{sec:hybridmodel}, we take three steps to develop the correction term needed in the muscle-torque hybrid model to accurately quantify exertion in above-shoulder interactions. In Study 1, we aim to obtain the shape of the curve of EMG-based fatigue indices at different vertical arm positions. We closely follow the experimental setup of the endurance arm-lifting task used in NICE~\cite{revisitingCE2023}, while introducing the following key changes:

\noindent\textbf{Angle granularity} - While NICE divided the vertical range of arm motion (30\textdegree\ - 150\textdegree) into five regions of 30\textdegree, in this study, we reduce the region size to 15\textdegree\ over a narrower range of arm motion (45\textdegree\ to 135\textdegree), resulting in a total of seven regions. Smaller regions better approximate a continuous distribution for vertical arm positions, enabling us to narrow the pivot angle of sudden changes.

\noindent\textbf{Maximum contraction duration} - We increase the maximum contraction duration from five to ten minutes. Previous works found the minimum contraction time to observe the linear relationship between EMG signals and the interaction time in bare-hand activity is 207-347s~\cite{naik2014applications}. Limiting the study duration to 5 minutes, as in \cite{revisitingCE2023}, prevents EMG signals from showing a clear increasing pattern and being used as fatigue indices.

\noindent\textbf{Natural arm weight} - We use self-weight only to make the study results valuable for bare-hand interactions. Meanwhile, \cite{revisitingCE2023} implemented varying arm weights to explore muscle fatigue of high-intensity tasks.

\noindent\textbf{Within-participant design} - Instead of the previous between-participant design, all participants finish the arm-lifting tasks in two separate sessions. A bigger sample size (n = 24) in the current study can better estimate the expected results in population.

\noindent\textbf{Participants} - We recruited 24 (compared to 12 in \cite{revisitingCE2023}) right-handed volunteers (12 female and 12 male - no participants identified as non-binary gender), aged 18-74 years, height 1.51-1.90 m, and weight 47-87 kg. Inclusion criteria for participants included: age between 18 and 75 years, being right-handed, reading and understanding English and numbers, not having a history of seizures, epilepsy or motion sickness, or any head/arm disability or impairment.

\subsection{Study Setup Details}

\paragraph{Task} The study task is side-lifting the right arm with no extra weight at the desired \textbf{Shoulder Angle}, the independent variable in Study 1, for up to 10 minutes. The \texttt{Shoulder\_Angle} refers to the angle between the participants' arms and torsos. The chosen angles in Study 1 are \textit{\textbf{45\textdegree, 60\textdegree, 75\textdegree, 90\textdegree, 105\textdegree, 120\textdegree}, }and\textit{ \textbf{135\textdegree}}. Each condition is performed only once, and participants have a 5-minute break between conditions. The order of the conditions is balanced using a Latin Square. Since the current study follows a within-participant design, we separate seven conditions into two sessions with over 48 hours in between to mitigate participants' fatigue (three conditions in one session and four conditions in another session). The maximum study duration is one hour, including the break. The total study duration for each session, including MVC collection, is approximately 90 minutes. Participants report their perceived fatigue through Borg CR10 ratings at a one-minute interval from the start till the end of the study.

\paragraph{Dependent Variables} The dependent variables used to assess muscle fatigue for each \texttt{Shoulder\_Angle} in the current study are:

\begin{itemize}[leftmargin=0pt]
    \item[] \textbf{\texttt{Duration}}:  Duration of the trial in seconds. A trial is concluded either after 10 minutes or if there are any failures in maintaining the arms at the specified \texttt{Shoulder\_Angle}. 

    \textbf{\texttt{Borg\_Slope}}: We collect the self-reported perceived fatigue (between 0-10) through Borg CR10 ratings every minute during the study and fit a linear regression model to all ratings to obtain the slope that shows the growing rate of subjective fatigue.

    \item[] The EMG-based fatigue indices ($\delta\%MVC\cdot s^{-1}$) for the investigated muscle groups are:    
    \textbf{\texttt{TR\_Slope}}: upper trapezius;
    \textbf{\texttt{MD\_Slope}}: middle deltoid;
    \textbf{\texttt{AD\_Slope}}: anterior deltoid;
    \textbf{\texttt{IF\_Slope}}: infraspinatus.
    
\end{itemize}

In total, we have $7\times24 = 168$ measurements for \texttt{Duration}, 
\texttt{Borg\_Slope}, \texttt{TR\_Slope} ($\delta\%MVC\cdot s^{-1}$), \texttt{MD\_Slope} ($\delta\%MVC\cdot s^{-1}$), \texttt{AD\_Slope} ($\delta\%MVC\cdot s^{-1}$), and \texttt{IF\_Slope} ($\delta\%MVC\cdot s^{-1}$).

\paragraph{Apparatus} The current study uses four EMG sensors (Trigno, Delsys Inc, Boston, MA) to collect muscle contraction in the Upper Trapezius (UT), Middle Deltoid (MD), Anterior Deltoid (AD), and Infraspinatus (IF) muscles, the primary contributors of shoulder movement. EMG sensors captured at two kHz were digitally integrated with the Vicon Nexus Motion Capturing software (v2.12, VICON, Oxford, UK) and synchronised with arm motion tracking. After obtaining the participants' maximum muscle capacity during the MVC collection, we placed 14 reflective markers on the participants' right arm, shoulder, and torso. The marker-based arm motion tracking allowed us to precisely monitor arm lifting and elbow extension during the study at 50 Hz using 10 VICON Vantage cameras.

\paragraph{Signal Processing} The maximum amplitudes of EMG of all four muscles from MVC collection were identified using Delsys EMG Acquisition (v4.7.9, Delsys, MA, USA). These maximum amplitudes were then utilized to normalize the EMG signals in the endurance study trials. Furthermore, the EMG recordings from the study conditions underwent bandpass filtering ([50-250] Hz, 4th order Butterworth) using Visual3D (v6.0, c-motion, MD, USA). We calculated EMG-based fatigue indices as dependent variables using custom-written scripts in Matlab (MathWorks, Natick, MA, USA). After extracting the time domain feature-RMS and down-sampling each signal to 50 Hz, we fitted a linear regression model to the signal and recorded the slope for further fatigue analysis. Moreover, we analysed Borg CR10 data using the slope of a linear regression model fitted to all ratings recorded during the study.

\subsection{Results: EMG-based Fatigue Indices under Different Shoulder Angles}
\label{sec:fatigueindice}

The current study implemented seven shoulder abduction angles in an endurance study to investigate the effect of vertical arm positions over fatigue measured in \texttt{Duration}, \texttt{Borg\_Slope}, and EMG-based fatigue indices (i.e. \texttt{TR\_Slope}, \texttt{MD\_Slope}, \texttt{AD\_Slope}, \texttt{IF\_Slope}). The findings are critical to understanding the mapping between fatigue and muscle contraction under different vertical arm positions.

Prior studies have found that Borg CR10 increases with the shoulder abduction angle between 30\textdegree--90\textdegree\ during 10s contractions~\cite{pincivero2010rpe} while the endurance time decreases with the shoulder abduction angles between 0\textdegree, 45\textdegree, and 90\textdegree~\cite{farooq2012effect}. However, the effects at angles above 90\textdegree\ for both short and long-duration contractions have not been fully explored.

\begin{figure}[h]
    \centering
    \includegraphics[width = \linewidth]{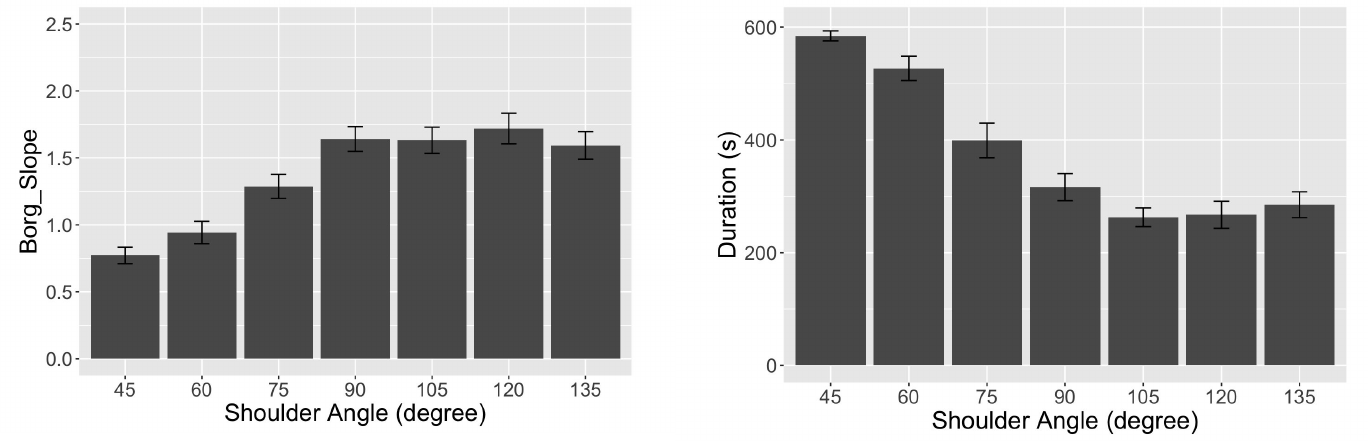}
    \caption{Mean values of \texttt{Borg\_Slope} (left) and \texttt{Duration} (right) under each \texttt{Shoulder\_Angle} in Study 1. Bars represent $\pm 1$ SE.}
    \label{fig:SubjectiveMeasures}
\end{figure}

As shown in Figure~\ref{fig:SubjectiveMeasures}, the patterns in \texttt{Borg\_Slope} and \texttt{Duration} during shoulder abduction below 90\textdegree\ are in line with previous studies~\cite{pincivero2010rpe,farooq2012effect} (i.e. \texttt{Borg\_Slope} increases with shoulder angle, while \texttt{Duration} decreases). In contrast, results above 90\textdegree\ deviate from this trend: in contrast to the symmetry predicted by prior torque-based models, we instead observe that both \texttt{Borg\_Slope} and \texttt{Duration} are relatively constant at \texttt{Shoulder\_Angle} between 90\textdegree\ and 135\textdegree\ .

\begin{figure}[h]
    \centering
    \includegraphics[width = \linewidth]{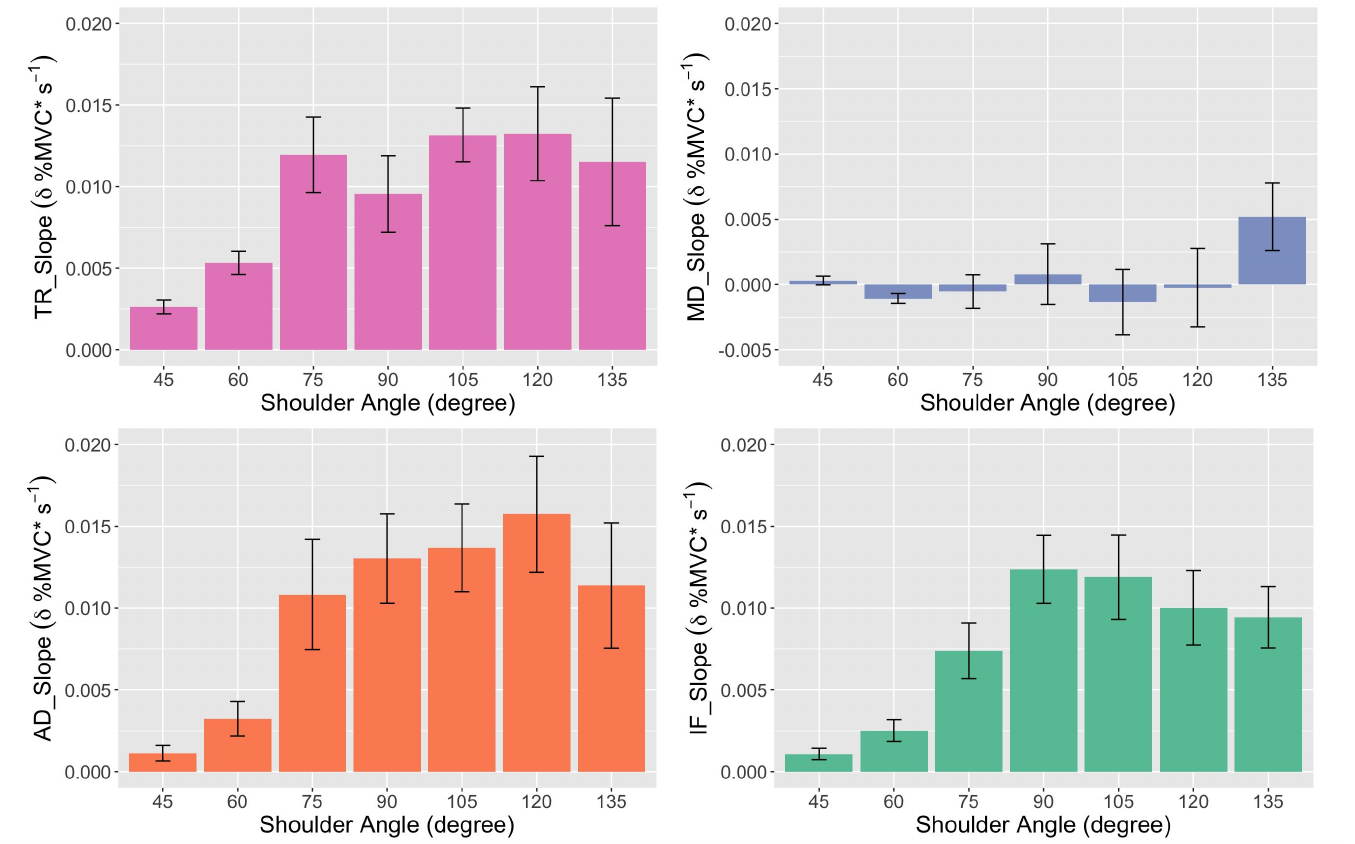}
    \caption{Mean values of \texttt{TR\_Slope} (top left), \texttt{MD\_Slope} (top right), \texttt{AD\_Slope} (bottom left), and \texttt{IF\_Slope} (bottom right) under each \texttt{Shoulder\_Angle} in Study 1. Bars represent $\pm 1$ SE.}
    \label{fig:EMGSlope}
\end{figure}

A similar pattern to \texttt{Borg\_Slope} and \texttt{Duration} can be seen in Figure~\ref{fig:EMGSlope}, which shows the individual response of each muscle. We can see that \texttt{TR\_Slope}, \texttt{AD\_Slope}, and \texttt{IF\_Slope} indicated that fatigue development was faster above 90\textdegree\ than below 90\textdegree\, while there was no significant difference between 90\textdegree\ and 135\textdegree. This observation is consistent with literature~\cite{wickham2010quantifying} where in a 5s shoulder abduction from 0\textdegree\ to 165\textdegree, the primary and secondary muscles of shoulder motion have the fastest growing rate of contraction between 0\textdegree\ and 90\textdegree, then reach the peak of contraction between 90\textdegree\ and 110\textdegree, and decrease after that with a slower rate than the rate at angles below 90\textdegree. Unlike the other muscles, MD exhibits positive slopes, indicating a failure to show fatigue, likely due to the compensatory effect of nearby muscles.

\begin{figure}
    \centering
    \includegraphics[width = 0.5\linewidth]{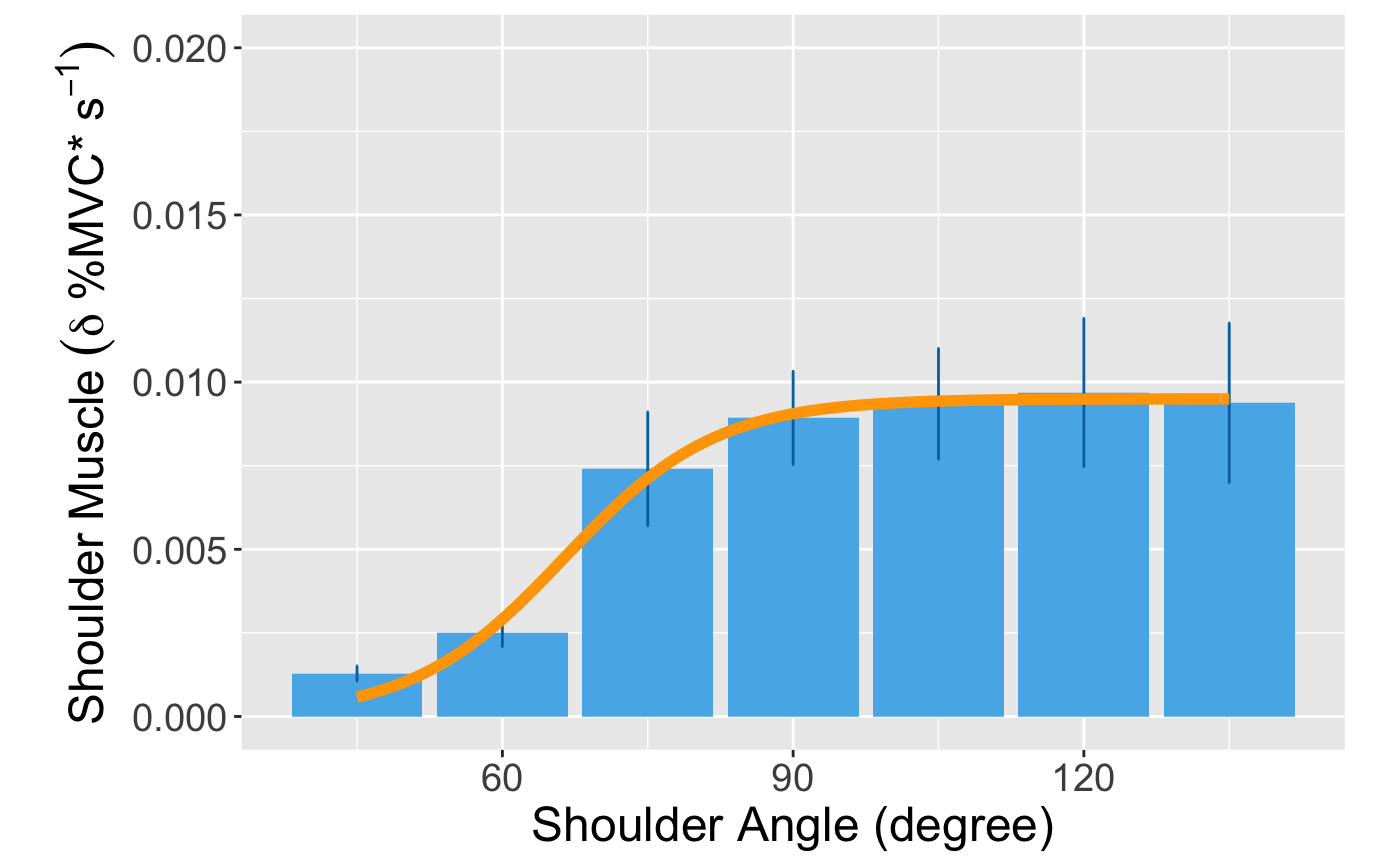}
    \caption{Mean values of the average of \texttt{TR\_Slope}, \texttt{MD\_Slope}, \texttt{AD\_Slope}, and \texttt{IF\_Slope} under each \texttt{Shoulder\_Angle} in Study 1. Bars represent $\pm 1$ SE. Orange line: the sigmoid curve estimated from the bar chart.}
    \label{fig:TruePattern}
\end{figure}

The consistent findings between the short-duration study~\cite{wickham2010quantifying} and the current long-duration study reveal the ground truth pattern of muscle fatigue development under different vertical arm positions. Therefore, both instantaneous physical exertion (i.e.\ shoulder torque measurements) and cumulative fatigue (i.e.\ existing fatigue models like CE and CF) should follow the pattern in Figure~\ref{fig:TruePattern} where ensemble averages (sum before average) for all four muscles were taken to allow joint-level comparison between different \texttt{Shoulder\_Angle}s. A sigmoid curve (Equation~(\ref{eq:sigmoid})) was estimated to fit the trend in Figure~\ref{fig:TruePattern}. The sigmoid function is used in Section~\ref{sec:model_development} to develop the correction term, $C(\theta)$, as visualised in Figure~\ref{fig:EMG_Torque}. The correction term will help the previous torque calculation to accurately account for the instantaneous exertion during above-shoulder interactions.

\begin{equation}
\label{eq:sigmoid}
     f(\theta) = \ \frac{0.0095}{1+\exp\left(\frac{\left(66.40-\theta\right)}{\ 7.83}\right)}\left\{0^{\circ}<\theta<180^{\circ}\right\}
 \end{equation}

\section{Study 2: Refinement of NICER Model with mid-air interaction tasks}

\label{sec:model_development}

Our findings from the endurance arm-lifting task in Study 1 establish the shape of the curve of muscle fatigue across different shoulder abduction angles. In Study 2, we use data from mid-air interaction tasks to determine the magnitude of shoulder torque and the curve's formula. Subsequently, we derive the necessary correction term needed for the muscle-torque hybrid model, which is previously explained in Section~\ref{sec:hybridmodel}. We then finalise the refined fatigue model, NICER, combining features of all previous models as per Table~\ref{tab:modelsummary}, at the end of the current section.

\paragraph{Task}
For our analysis, we use data from an immersive point selection task within 3D scatterplots (discussed in Section~\ref{sec:StudyDataSet}). This task allows participants to perform full-range mid-air gestures without constraining shoulder abduction and elbow extension. We combine data from two different interaction methods designed to explore distant selection in the original study. The first (called Linear Gain in the original study) is a variation of traditional `virtual hand extension' methods (e.g. Go-Go~\cite{gogo1996}) in VR, in this case connected by a telescopic arm. The second (called Haptic Portal) is a novel implementation that allows the user to reach through a pair of portals to reach a robotic arm-assisted haptic target. (See Appendix~\ref{sec:AppendixDataSet} for further details.) For the purposes of our model refinement, we do not differentiate these methods as a study variable. We combine the data from both interaction methods to provide a high variation in task time and arm motion.

\paragraph{Independent Variables}
As described in Section~\ref{sec:hybridmodel}, we aim to use data from a representative constraint-free mid-air interaction task to investigate the distribution of shoulder torque under varying shoulder angles and later use it to develop the refined model formulation. 
Therefore, the sole independent variable is the continuous real-time \textbf{Shoulder Angle} (\textit{\textbf{0\textdegree}} $\leq \theta \leq$  \textit{\textbf{180\textdegree}}) performed during the interaction task.

\paragraph{Dependent Variables}
To understand the distribution of shoulder torque during mid-air gestures of extended arms, we collect measures of \textbf{\texttt{Shoulder Torque}}, defined as follows:
The real-time shoulder torque in Newton–meter measured in the mid-air selection task when elbow flex angle $\leq 35$\textdegree. Detailed calculation is in \cite{Hincapie-Ramos_CE_2014}.

 \paragraph{Apparatus}
 The arm movement was tracked by a Vicon motion-capture system. Additionally, a MANUS glove was used to track the precise hand gestures when grabbing the virtual target in VR. 

\paragraph{Participants}
The study included 24 right-handed volunteers (7 female and 17 male - no participants identified as non-binary gender), aged 19-35 years, height 1.58-1.82 m, and weight 43-90 kg. Only seven participants had no prior experience with VR. It is worth noticing that participants involved in Studies 2 and 3 do not overlap. One participant appeared in Studies 1 and 2.

\subsection{Correction Term to Revise the Torque Calculation in Dynamic 3D Interactions}
\label{sec:correctionterm}

\begin{figure}[h]
    \centering
    \includegraphics[width = \linewidth]{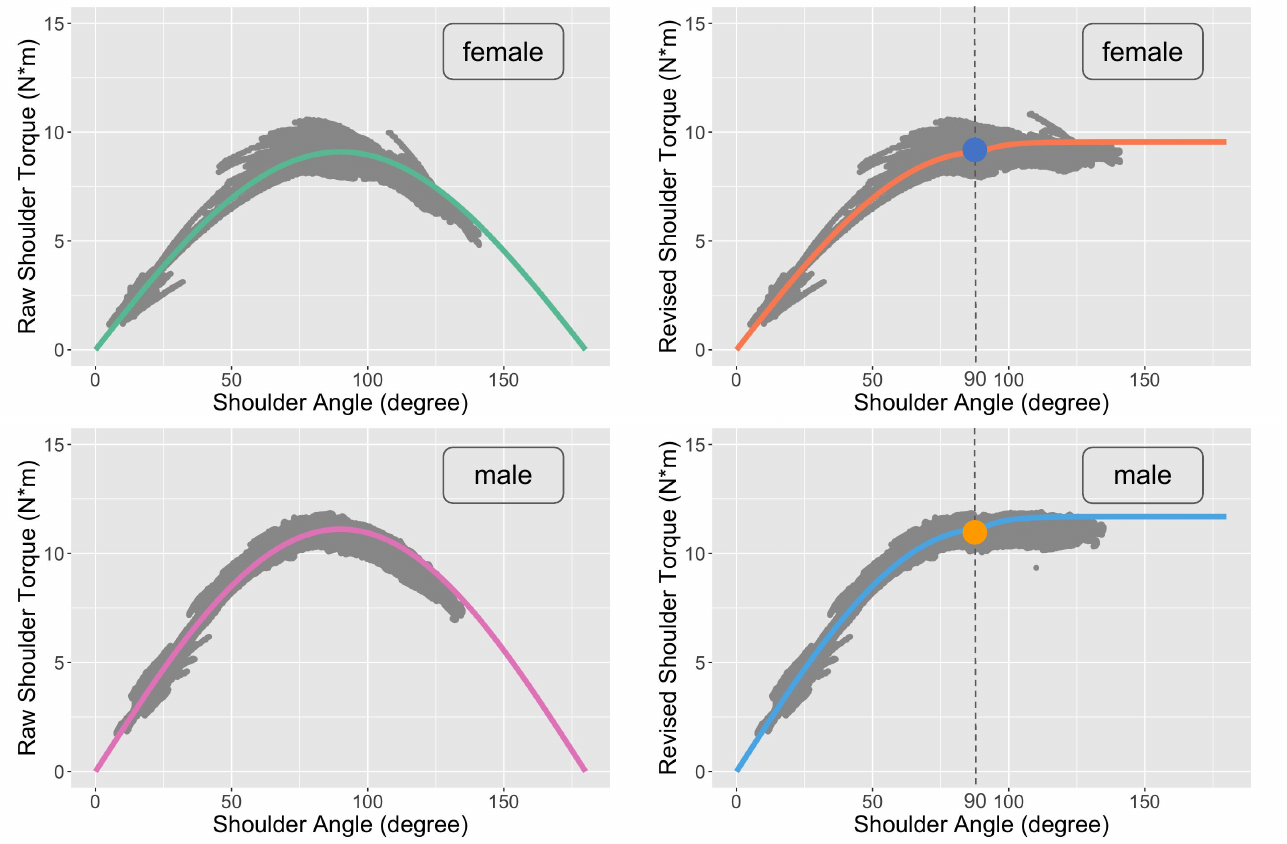}
    \caption{Left: The scatter plot in grey is the torque distribution of the extended arm interaction in the mid-air selection tasks. The green curve depicts $t_{female}$ in Equation (\ref{eq:femaletorque}) while the pink curve shows $t_{male}$ in Equation (\ref{eq:maletorque}). Right: The scatter plot in grey is the revised torque distribution of the extended arm interaction in the mid-air selection tasks. Only torque collected when shoulder angle $>$ 90\textdegree\ was revised by adding the correction term. The orange curve is the combination of the green curve and $C_{female}(\theta)$ in Equation (\ref{eq:femalecorrectionterm}) and the blue curve is the pink curve combined with $C_{male}(\theta)$ in Equation (\ref{eq:malecorrectionterm}).}
    \label{fig:s2_revisedtorque}
\end{figure}

The earlier exploration of CE~\cite{revisitingCE2023} revealed a limitation in using only shoulder torque to account for exertion above the shoulder when the elbow is extended (i.e. elbow flex angle $\leq 35$\textdegree). This underscores the need for a muscle-torque hybrid approach to accurately quantify instantaneous exertion. After establishing the shape of the fatigue curve (Equation~(\ref{eq:sigmoid})) in the above-shoulder activity in Study 1, we now use data collected in Study 2 to get the additional term to match torque measures with the curve. To determine the required scaling to make the decreasing shoulder torque increase to a similar level when the shoulder angle $\theta >$ 90\textdegree, we first fitted a sine function: $t(\theta)$ to analyze the raw torque distribution during the extended-arm mid-air selection task for female and male participants. This is detailed in Equations~(\ref{eq:femaletorque}) and (\ref{eq:maletorque}) (see Figure~\ref{fig:s2_revisedtorque}-left). 

\begin{equation}
\label{eq:femaletorque}
    t_{female}(\theta) =\ \frac{\sin\left(\theta\cdot\frac{2\pi}{360}\right)}{0.11}\left\{0^{\circ} <\theta<180^{\circ} \right\}
\end{equation}

\begin{equation}
\label{eq:maletorque}
    t_{male}(\theta) =\ \frac{\sin\left(\theta\cdot\frac{2\pi}{360}\right)}{0.09}\left\{0^{\circ}<\theta<180^{\circ}\right\}
\end{equation}

We then obtained the scale-up factor $\beta$ of 1005 for female participants and 1230 for male participants by setting the ground truth curve $f(\theta)$ to the same level as the raw shoulder torque $t(\theta)$ at 90\textdegree, i.e.  $f(90^{\circ}) = t(90^{\circ})$ (see Figure~\ref{fig:s2_revisedtorque}-right). Subsequently, we defined the correction term $C$ as the difference between $\beta\cdot f(\theta)$ and $t(\theta)$ for $\theta > 90^{\circ}$ in Equation~(\ref{eq:femalecorrectionterm}) and (\ref{eq:malecorrectionterm}). As shown in Figure~\ref{fig:s2_revisedtorque}-right, the shoulder torque measured above $90\textdegree$ has been successfully revised and now aligns with the pattern in Figure~\ref{fig:TruePattern}. Moving forward, we will integrate $C_{female}(\theta)$ and $C_{male}(\theta)$ with the original torque calculation from CE to revise the shoulder torque value when the elbow is extended and the interaction occurs above the shoulder.

\begin{equation}
\label{eq:femalecorrectionterm}
    C_{female}(\theta) =\ \frac{0.0095\cdot1005}{1+\exp\left(\frac{\left(66.40-\theta\right)}{\ 7.83}\right)}-\frac{\sin\left(\theta\cdot\frac{2\pi}{360}\right)}{0.11}\left\{90^{\circ}<\theta<180^{\circ}\right\}
\end{equation}

\begin{equation}
\label{eq:malecorrectionterm}
    C_{male}(\theta) =\ \frac{0.0095\cdot1230}{1+\exp\left(\frac{\left(66.40-\theta\right)}{\ 7.83}\right)}-\frac{\sin\left(\theta\cdot\frac{2\pi}{360}\right)}{0.09}\left\{90^{\circ}<\theta<180^{\circ}\right\}
\end{equation}

The current study retrieved the correction term from arm movement data collected in a constraint-free interaction task. This allows the refined model to work with any bare-hand interactions without requiring an additional model training process. It is worth noting that our approach is different from the supervised training approach of CF. While CF obtains model parameters by minimising the error between fatigue predictions and subjective fatigue ratings, our study develops the correction term based on the distribution of shoulder torque in bare-hand arm movements.

\subsection{Finalising the NICER Model Formulation}
\label{sec:NICERformula}

To finalise the comprehensive refined fatigue model, NICER, we took the following steps. We updated the ET function in CE to a shoulder joint-specific ET function built on the grand average ET of the literature ~\cite{frey2010endurance}. Combining with the correction term $C(\theta)$ developed in Section~\ref{sec:correctionterm}, $Max\_Torque$ from Chaffin's maximum strength model in Appendix~\ref{sec:chaffinmodel}, and the recovery factor in Section~\ref{sec:recoveryfactor}, we finalise NICER formulation in Equation~(\ref{eq:eqnicer_active}) and (\ref{eq:eqnicer_rest}).

\begin{equation}
 \label{eq:eqnicer_active}
     NICER_{active} = \frac{Duration\cdot (\frac{Torque + C(\theta)}{Max\_Torque}*100)^{1.83} \cdot 0.000218}{14.86} * 100
\end{equation}

\begin{equation}
 \label{eq:eqnicer_rest}
     NICER_{rest} = NICER_{active} \cdot \exp ^{-0.04 \cdot \delta t}
\end{equation}

\section{Study 3: NICER Model Validation}
\label{sec:model_evaluation}

The comprehensive NICER model that includes all desired features summarised in Table~\ref{tab:modelsummary} is finalised in the last section. The goals of Study 3 are twofold: first, we compare NICER with subjective interaction measures to see whether NICER can distinguish the low-fatigue condition from the high-fatigue condition by the estimated fatigue levels. Second, we evaluate NICER and CF (2017 and 2023 versions) in a study of varying interaction durations to assess their performance in a complex interaction task.

We exclude NICE from this comparison as it has yet to be validated in interaction tasks. In our experiments, CE fails to make fatigue predictions for the current mid-air selection tasks due to its limitations in low-intensity exertion. Therefore, we did not consider CE in the following analysis. \chdeleted{Furthermore, as the latest CF model is not readily accessible, we can only discuss the results of the original CF model. However, prior work found no significant difference between the original CF and its recently revised variant before adding Chaffin's model.}

\paragraph{Task}
For this evaluation, we use two interaction methods from our external dataset. As discussed in Section~\ref{sec:StudyDataSet}, we chose these methods for our evaluation because they were previously shown to have significantly different subjective fatigue scores \footnote{The observation is due to differences in selection difficulty.  This also resulted in apparent differences in interaction time, however, the difference in means was not statistically significant. See original paper~\cite{magicportal2023} for details.}. For our evaluation purposes, we aptly label these \textit{High\_Fatigue} and \textit{Low\_Fatigue}.

The \textit{High\_Fatigue} method (called Adaptive Gain in the original study) is a variation of the Linear Gain method described in Section~\ref{sec:model_development} (with an adaptive gain function in place of a linear one. See Appendix~\ref{sec:AppendixDataSet} for implementation details). Likewise, the \textit{Low\_Fatigue} method (called Portals) is a variant of the previous Haptic Portal method (without the robot-assisted haptic feature). 

\begin{figure}
    \centering
    \includegraphics[width = \linewidth]{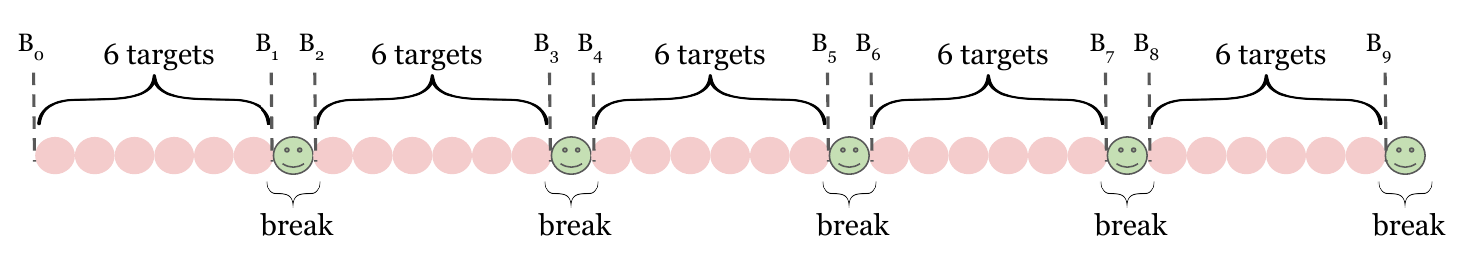}
    \caption{Study procedure of the mid-air selection task in the external dataset~\cite{magicportal2023}. Each break lasts 15s, and $B_{i}$ represents the $i$th Borg CR10 rating collected during the study.}
    \label{fig:S2_workflow}
\end{figure}

The target selections were divided into several blocks. Each block included six target selections, after which participants took a 15s break to report their perceived fatigue through Borg CR10. We call every set of six targets one "block" in the result section, and for every block, we report two Borg CR10 values, one collected at the beginning and the one at the end of the block.  Participants completed one full set of tasks (30 trials/5 blocks) with each interaction method. (see Figure~\ref{fig:S2_workflow}).

\paragraph{Independent Variables}
{As described in Section~\ref{sec:proposed studies}, we evaluate different fatigue models' strengths for better guiding interaction designs. For this evaluation, we use one independent variable: \textbf{Interaction Technique} with two levels: \textit{High\_Fatigue} and \textit{Low\_Fatigue}.

\paragraph{Dependent Variables}
In the current study design, we aim to control the target placement and hand trajectory between conditions to discover the influence of varying study durations on the dependent variables below:

\begin{itemize}[leftmargin=0pt]
    \item[] \textbf{\texttt{Borg CR10}}: Self-reported perceived fatigue on a 0--10 scale (0: nothing at all, 10: extremely strong). 
    \item[] Target-wise fatigue predictions at the beginning of the study and after each successful target selection: \textbf{\texttt{CF\_2017}} (\%), \textbf{\texttt{CF\_2023}} (\%), and \textbf{\texttt{NICER}} (\%). 
    \item[] Block-wise fatigue predictions generated when participants report the Borg score after every six targets: \textbf{\texttt{CF\_2017}} (\%), \textbf{\texttt{CF\_2023}} (\%), and \textbf{\texttt{NICER}} (\%).
\end{itemize}

For each condition, we have $10\times12 = 120$ measures for \texttt{Borg CR10}, block-wise \texttt{CF} and \texttt{NICER}, and $31\times12 = 372$ measures of target-wise \texttt{CF\_2017}, \texttt{CF\_2023}, and \texttt{NICER}.

\paragraph{Participants}
Data were collected from 12 right-handed volunteers (5 female and 7 male - no participants identified as non-binary gender), aged 18-35 years, height 1.53-1.80 m, and weight 54-82 kg. Only two participants had no prior experience using VR. It is important to note that the participants in Studies 2 and 3 do not overlap. One participant appeared in Studies 1 and 3.

\subsection{Results}
\label{sec:study_result}

We used a pair-wise $t$-test \chreplaced{with an alternative hypothesis that the mean of fatigue predictions under the \textit{High\_Fatigue} condition are greater than the mean of predictions under the \textit{Low\_Fatigue} condition to compare}{two conditions for} all dependent variables. The detailed results are reported below and summarized in Table~\ref{tab:paired-t-test}.

\begin{figure}[h]
    \centering
    \includegraphics[width = \linewidth]{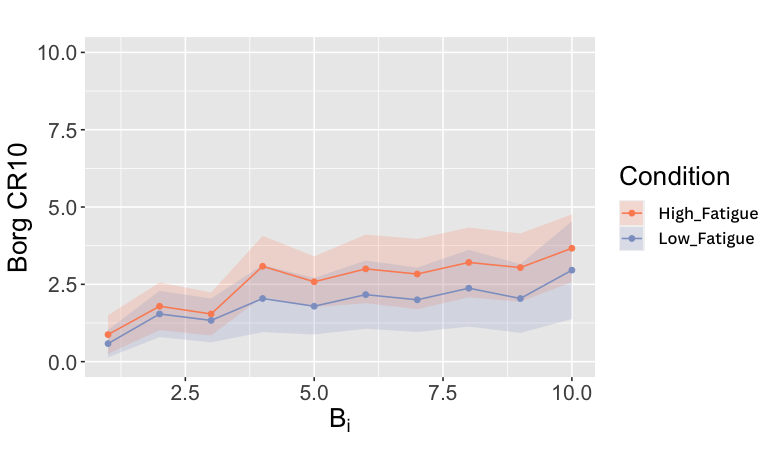}
    \caption{Mean value of the block-wise results of \texttt{Borg CR10} between conditions. The width of the ribbon represents $95\%$ CI~\cite{magicportal2023}.}
    \label{fig:s2_borg_measures}
\end{figure}

\paragraph{Borg CR10}
\noindent The \textit{Low\_Fatigue} condition achieved a significantly lower \texttt{Borg CR10} than the \textit{High\_Fatigue} condition ($t_{119} = 6.264$, $p < .001$).
See Figure~\ref{fig:s2_borg_measures}.

\paragraph{CF\_2017}
\noindent The fatigue predictions estimated by \texttt{CF\_2017} \chreplaced{consider \textit{Low\_Fatigue} conditions as more fatiguing than \textit{High\_Fatigue} conditions}{for \textit{Low\_Fatigue} and \textit{High\_Fatigue} conditions were too similar to differentiate} in both target-wise ($t_{371} = -1.362$,  \chreplaced{$p =.913$}{.174}) and block-wise ($t_{119} = -1.028$, \chreplaced{$p =.847$}{.306}) results. See Figure~\ref{fig:s2_NICER_CF}-top.

\paragraph{CF\_2023}
\noindent \chadded{The fatigue predictions estimated by \texttt{CF\_2023} consider \textit{Low\_Fatigue} conditions as more fatiguing than \textit{High\_Fatigue} conditions in both target-wise ($t_{371} = -4.868$, $p = 1$) and block-wise ($t_{119} = -2.709$, $p = .996$) results. See Figure~\ref{fig:s2_NICER_CF}-middle.}

\paragraph{NICER}
\noindent From Figure~\ref{fig:s2_NICER_CF}-bottom, we can see that the objective fatigue measure \texttt{NICER} considered the \textit{High\_Fatigue} condition as significantly more fatiguing than the \textit{Low\_Fatigue} condition in both target-wise ($t_{371} = 8.147$, $p < .001$) and block-wise ($t_{119} = 4.503$, $p < .001$) results.

\begin{figure}[h]
    \centering
    \includegraphics[width = \linewidth]{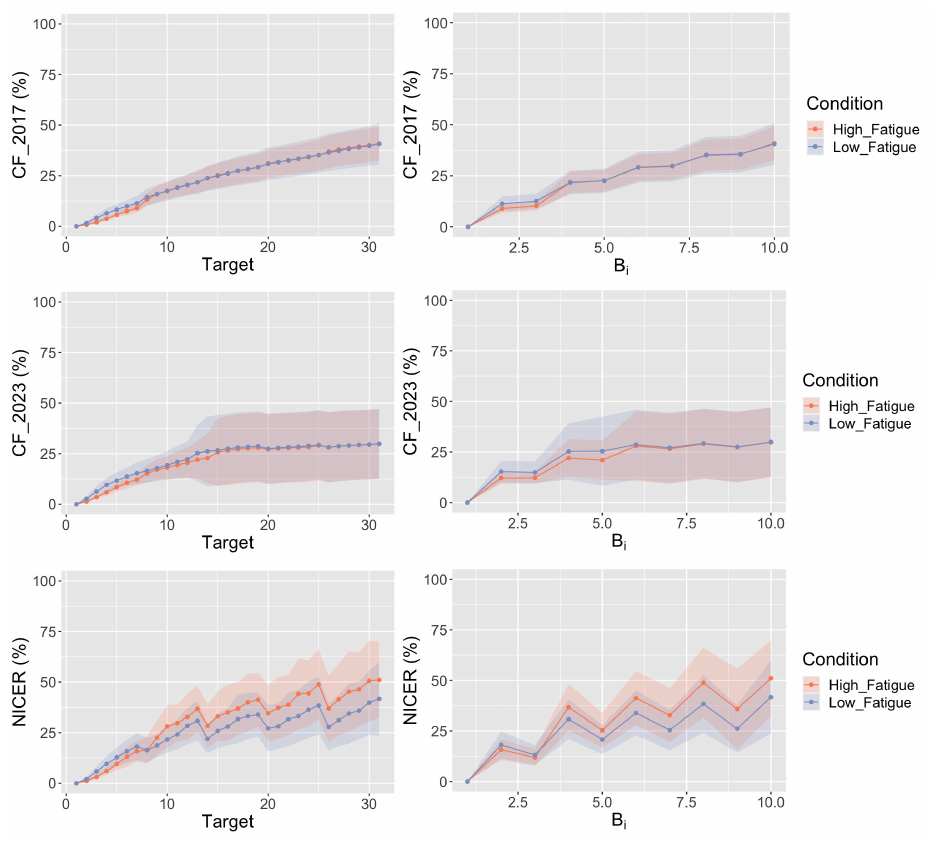}
    \caption{Top: Mean value of the target-wise ($Target$) result of CF\_2017 between conditions (left); Mean value of the block-wise ($B_{i}$) results of CF\_2017 between conditions (right). \chadded{Middle: Mean value of the target-wise results of CF\_2023 between conditions (left); Mean value of the block-wise results of CF\_2023 between conditions (right).} Bottom: Mean value of the target-wise results of NICER between conditions (left); Mean value of the block-wise results of NICER between conditions (right). The width of the ribbon represents $95\%$ CI.}
    \label{fig:s2_NICER_CF}
\end{figure}

\begin{table}[h]
\caption{Results of the pairwise $t$-test of average predictions under the Alternative Hypothesis: $\mu_{High\_Fatigue}>\mu_{Low\_Fatigue}$}
\begin{tabular}{|c|c|c|}
\hline
Paired $t$-test & Target-wise & Block-wise \\ \hline
Borg CR10 & NA & $t_{119}$ = 6.264, \textbf{p < .001} \\ \hline
CE & NA & NA \\ \hline
\chadded{CF\_2017} & $t_{371}$ = -1.362, p = \chreplaced{.913}{.174} & $t_{119}$ = -1.028, p = \chreplaced{.847}{.306} \\ \hline
\chadded{CF\_2023} & \chadded{$t_{371}$ = -4.868, p = 1} & \chadded{$t_{119}$ = -2.709, p = .996} \\ \hline
NICER & $t_{371}$ = 8.147, \textbf{p < .001} & $t_{119}$ = 4.503, \textbf{p < .001} \\ \hline
\end{tabular}
\label{tab:paired-t-test}
\end{table}

\chadded{We calculated the Pearson Correlation Coefficient $\rho$ to assess the similarity between fatigue model predictions (i.e. NICER and CF scores) and subjective fatigue ground truth (i.e. \texttt{Borg CR10}) in comparing \textit{Low\_Fatigue} and \textit{High\_Fatigue} conditions. NICER achieved the highest $\rho$ with \texttt{Borg CR10} in the \textit{Low\_Fatigue} condition ($\rho = 0.976$) and in the \textit{High\_Fatigue} condition ($\rho = 0.978$). Meanwhile, CF\_2023 showed a higher correlation ($\rho = 0.966$) than CF\_2017 ($\rho = 0.954$) in the \textit{High\_Fatigue} condition but a lower correlation ($\rho = 0.923$) than  CF\_2017 ($\rho = 0.940$) in the \textit{Low\_Fatigue} condition. (see Table~\ref{tab:correlationsoeeficient}). The above results show that NICER makes more accurate fatigue predictions than CF\_2017 and CF\_2023.}

\begin{table}[]
\caption{Pearson Correlation Coefficients $\rho$ between Borg CR10 (average) and CE (average), block-wise CF\_2017 (average), \chadded{block-wise CF\_2023 (average),} and block-wise NICER (average), as well as their corresponding 95\% CI.}
\begin{tabular}{|c|c|c|}
\hline
$\rho$ & High\_Fatigue & Low\_Fatigue \\ \hline
CE & NA & NA \\ \hline
CF\_2017 & 0.954 [0.8123, 0.9894] & 0.940 [0.7604, 0.9860] \\ \hline
\chadded{CF\_2023 } & \chadded{0.966 [0.8568, 0.9921]} & \chadded{0.923 [0.7005, 0.9820]}\\ \hline
NICER & \textbf{0.978} [0.9057, 0.9950] & \textbf{0.976} [0.9003, 0.9945] \\ \hline
\end{tabular}
\label{tab:correlationsoeeficient}
\end{table}

\subsection{Discussion: Is NICER a Reliable Analytical Tool?}
\label{sec:NICER_CF}

 To determine whether NICER provides a reliable tool for the analysis of potential interaction methods, we consider the three criteria proposed in Section~\ref{sec:proposed studies}.

\paragraph{Criterion 1: Can fatigue models generalise to different interaction designs?}
A key attribute of a reliable tool is to generalise to different interactions, including variations in intensity, range of motion and duration. Our evaluation tests the ability of the models to differentiate between two different interaction methods with significant differences in subjective fatigue scores (see Table~\ref{tab:paired-t-test}). Results in Section~\ref{sec:study_result} show that NICER is able to differentiate between these tasks at both target-wise and pair-wise levels, whereas \chreplaced{CF\_2017 and CF\_2023}{CF} are not. This confirms that NICER has a stronger power in distinguishing interaction techniques based on the predicted fatigue than \chreplaced{CF\_2017 and CF\_2023}{CF}.

\paragraph{Criterion 2: Can fatigue models reflect fatigue recovery during breaks in activity?}
In the mid-air selection tasks, \texttt{Borg CR10} was collected once before the break and once after the break to directly observe varying subjective perceived fatigue during rest periods in the study trial. Generally, participants perceived increasing fatigue during task blocks but received noticeable fatigue recovery after the 15s break.

As can be seen in Figure~\ref{fig:s2_NICER_CF}, NICER shows a similar fatigue variation as \texttt{Borg CR10}, where the predictions increased with the target blocks and decreased after the break started. Compared with the original CE, which considers the estimated fatigue level as constant during the interaction break, NICER improves significantly in quantifying fatigue recovery by showing a clear drop between the active task and the rest periods. On the other hand, \chreplaced{CF\_2017}{CF} shows a continuously increasing pattern during the study and fails to reflect the rest period in reducing the estimated fatigue levels. \chadded{CF\_2023 performs better than CF\_2017 by showing reduced fatigue predictions after nearly every break. However, the decrease is only subtle throughout the entire study.}

\chdeleted{We calculated the Pearson Correlation Coefficient $\rho$ to assess the similarity between fatigue model predictions (i.e. NICER and CF scores) and subjective fatigue ground truth (i.e. \texttt{Borg CR10}) in comparing \textit{Low\_Fatigue} and \textit{High\_Fatigue} conditions. NICER achieved a higher $\rho$ with \texttt{Borg CR10} in the \textit{Low\_Fatigue} condition ($\rho = 0.976$) and in the \textit{High\_Fatigue} condition ($\rho = 0.978$) than CF with \texttt{Borg CR10} in the \textit{Low\_Fatigue} condition ($\rho = 0.940$) and in the \textit{High\_Fatigue} condition ($\rho = 0.954$) (see Table~\ref{tab:correlationsoeeficient}).}

\paragraph{Criterion 3: Can fatigue models accurately predict fatigue levels?}
As discussed in Section~\ref{sec:proposed studies}, NICER and CF provide different predictions (endurance time versus perceived fatigue). Whereas CF units can be directly converted to Borg units using a conversion factor of $0.0875$, a similar conversion of NICER has not been validated~\cite{jang2017modeling}. 

To evaluate their predictive capability, we compare the overall range of NICER and CF to the range of reported Borg scores. Overall, the results of CF\_2017 and CF\_2023 (range between $30\%-50\%$) have a similar magnitude as \texttt{Borg CR10} (range between $13\%-48\%$ of the maximum value 10) while NICER (range between $24\%-69\%$) has a higher magnitude that will potentially overestimate the perceived fatigue. This observation is expected since model parameters of CF were optimised to fit \texttt{Borg CR10} scores. We will discuss this further in Section~\ref{sec:limitations}.

\subsection{Summary}
In comparing fatigue model predictions with the subjective measures reported in Section~\ref{sec:study_result}, we found varying strengths and weaknesses of NICER and CF. We found that NICER can differentiate between two interaction methods as similarly was observed in the objective measures. This indicates that NICER possesses a greater ability to generalise to different tasks than CF. The inability of CF to differentiate these interaction methods may be due to overfitting Borg CR10 scores in its supervised training approach.

We also found that NICER shows a greater variation between periods of activity and recovery than CF for the interactions investigated in our study. A visual inspection of the predictions in Figure~\ref{fig:s2_NICER_CF}, in comparison with the subjective scores in Figure~\ref{fig:s2_borg_measures}, shows that the recovery approach in NICER may be too aggressive, whereas CF fails to drop after every break. Additionally, Pearson Correlation Coefficients in Table~\ref{tab:correlationsoeeficient} have a stronger fit for NICER than CF for both interaction methods, however, overlapping confidence intervals indicate that these differences may not be significant. Overall, these results indicate that NICER is able to better predict fatigue recovery than previous models.

When looking at the predicted range of fatigue levels induced by different activities, we found that \chreplaced{f}{NICER may overestimate fatigue in comparison to the subjective measures. F}atigue predictions by CF achieved a smaller error to the subjective measures than NICER, which implies that NICER may overestimate the subjective perceived fatigue.

\section{Applications and Future Work}
The study results of the mid-air selection task in Section~\ref{sec:model_evaluation} show that our proposed refined model NICER can effectively distinguish the study conditions and make accurate fatigue predictions that capture fatigue recovery during breaks. We are confident that from now on, NICER can serve as the objective fatigue measurement and guide future interaction design.

\subsection{Applications of NICER}
\label{sec:applications}
\chadded{NICER has numerous applications due to its comprehensiveness. We anticipate its integration into wearable and mobile interaction design, the Internet of Things, mobile VR, AR, and Mixed Reality (MR) systems, character simulation~\cite{cheema2023discovering}, and mid-air haptics~\cite{long2014rendering}, among other exciting applications.}

\chadded{A straightforward use of NICER is as an objective metric to optimise the design of graphical user interfaces~\cite{oulasvirta2020combinatorial} in order to develop a safe and ergonomic-friendly user experience. NICER can assist designers in identifying the most comfortable positions to place static 3D UIs~\cite{yu2019modeling} by adapting the current arm gestures and the interaction time. For example, the 3D object position can be re-targeted to reduce arm movement in VR object retrieval~\cite{jiang2023commonsense} to enable an extended engagement. Similarly, interaction techniques can be made adaptive by shifting input modality, for instance, from gesture to speech.}

\subsection{Limitations and Future Work}
\label{sec:limitations}

First, our investigation of muscle fatigue development in Section~\ref{sec:study1} was limited to the vertical plane only (i.e. 0\textdegree\ in the horizontal plane). This was chosen to focus on muscles that are the primary contributors to vertical shoulder movements and to prevent interference from muscles involved in horizontal movements. However, since shoulder torque remains at the same level when the shoulder is at a fixed vertical position, torque may fail to capture the physical exertion that varies between different horizontal positions. A similar concern was raised in applying CE in 3D UI design~\cite{Belo_XRgonomics_2021}, where participants found the interaction at their left sides was more fatiguing than the right side when interacting with their right arms. Future work could follow a similar design in the endurance study in Section~\ref{sec:study1} but with several arm angles at the horizontal plane to investigate muscle fatigue development under different horizontal arm positions. The empirical comparison between shoulder torque and EMG-based fatigue indices may contribute to another correction term for interaction above 90\textdegree\ in the horizontal plane. Moreover, results combined with our findings in the vertical plane (Section~\ref{sec:fatigueindice}) will lead to the complete picture of muscle fatigue development in the full range of shoulder motion. As such, future studies can apply these results in biomechanical simulations to estimate the fatigue indices for a given arm position, similar to the prior study~\cite{bachynskyi2015informing} on estimating the muscle activation cost based on the coordinate in the interaction volume, \chadded{and the work on modelling neck muscle fatigue in XR~\cite{zhang2023toward}}.

The current study evaluated the model performance by comparing fatigue predictions with subjective interaction measures. Though results confirm the effectiveness of using NICER in analysing the difference between the conditions chosen for this work, further study is needed to determine whether the models generalise to a greater variety of interactions \chadded{(e.g., dynamic task intensities and varying recovery times)} in real-world settings. 

It remains unclear how close the fatigue prediction of NICER is to the ground truth subjective sores. Previous evaluation done in CF~\cite{jang2017modeling,villanueva2023advanced} applied a linear conversion factor to assess the root mean squared errors (RMSE) between model predictions and Borg CR10. However, such a conversion has not been empirically validated. Since we cannot assume that NICER and Borg CR10 use the same scale, we instead applied Pearson Correlation to compare the general patterns. A potential future direction is to explore a feasible way to quantify the distance between model predictions of endurance time and subjective fatigue measures.

\chreplaced{In the model development of NICER, a young group of participants contributed to the study. Furthermore, the current model formulation only considers the variation of participants between sexes.}{In the original CE implementation, only a sex-based factor was taken to consider the variation of participants. Therefore, our NICER model has the same limitation.} However, physical effort is sometimes affected by gender~\cite{lindle1997age}, age~\cite{lindle1997age}, and height~\cite{ford2000gender}. \chadded{We encourage future studies to investigate the effect of age on fatigue models.} We plan to represent these human factors from a high-level perspective by introducing the upper limb length as one of the model parameters in future implementation. CF uses participants' arm length as one of the model inputs. However, a time-consuming pre-study calibration is needed to take the measurement. A future improvement can be using the inverse kinematic toolkit to estimate the arm length before running the model.

Lastly, literature, including the current study, only considers single right-arm interactions when developing the fatigue model. However, inputs from bimanual interaction and full-body movement are becoming popular in 3D gaming and multi-touch large-screen interactions. Future work can generalise the takeaway of the current literature and extend the improved models to consider body segments rather than shoulder and arm. 

\section{Conclusion}
In this paper, we introduce a refined model to quantify shoulder fatigue in mid-air interaction, called \textit{NICER}, which is short for ``New and Improved Consumed Endurance and Recovery Metric". NICER is a ready-to-use, lightweight but comprehensive model that can: (1) predict the maximum duration of interaction from arm position; (2) work with interactions of varying physical intensities; (3) capture the extra exertion of above-shoulder arm movement; (4) consider gesture-based maximum strength; and (5) reflect fatigue recovery after rest periods. With the latest head-mounted display (HMD) devices, the necessary inputs for NICER can be easily retrieved from real-time hand-tracking. Thus, NICER is ready to be used in immersive interaction design. 

In Study 1, we investigated muscle fatigue at different shoulder abduction angles to obtain the curve of objective muscle fatigue. We used arm movement data collected from a mid-air interaction task with no constraints on shoulder positions and elbow extension to finalise the refined model formulation in Study 2. Finally, in Study 3, we evaluated the model performance as a design analytical tool in a mid-air selection task of varying durations. The promising study results show that NICER can objectively distinguish the interaction designs of different perceived fatigue and accurately reflect the decreasing fatigue after the break. A stronger correlation with the subjective ground truth of perceived fatigue than the previous model (\cite{jang2017modeling, villanueva2023advanced}) confirms that NICER is a reliable fatigue measurement to assist future interaction design.

\begin{acks}
We thank Sujin Jang and Ana Villanueva for providing access to the source code for their CF model. We also thank our study participants for their time and exertion, as well as anonymous reviewers for their insightful feedback.
\end{acks}

\bibliographystyle{ACM-Reference-Format}
\bibliography{NICER_SIGGRAPH24}

\appendix
\section*{Appendix}
\renewcommand{\thesubsection}{\Alph{subsection}}

\section{Chaffin’s Model of Maximum Strength Estimation}
\label{sec:chaffinmodel}

\chadded{Chaffin's model outputs the current $Max\_Torque$ $(N\cdot m)$ at the shoulder joint based on the performed arm gestures represented in the elbow extension angle ($\alpha_{e}$) and shoulder abduction angle ($\alpha_{s}$) while considering the difference of physical capacity between gender by factor $G~(G_{female} = 0.1495, G_{male} = 0.2845)$. The full model formula is shown in Equation~\ref{eq:chaffinmodel}.}

\begin{equation}
\label{eq:chaffinmodel}
    Max\_Torque = (227.338 + 0.525 \cdot \alpha_{e} - 0.296 \cdot \alpha_{s})\cdot G
\end{equation}

\section{Summary of External Dataset}
\label{sec:AppendixDataSet}
As mentioned in Section~\ref{sec:StudyDataSet}, a target selection task is chosen in Studies 2 and 3 to finalise and evaluate NICER. This section provides a summary of the study task and interaction method implemented in the original study~\cite{magicportal2023}. We refer readers to the original paper for complete details. 

\subsection{Interaction Task: Distant Object Manipulation}
\label{sec:AppendixTask}

Participants are asked to interact with distant objects using their hands after being introduced to the study environment in a virtual room in VR. In detail, participants must successfully reach and select a set of 30 randomly ordered targets, one by one, from a cluster of data points in each trial, using a specific interaction method, with an extended virtual arm, controlled by their right arms, as visualised in Figure~\ref{fig:s2setup}-right.\chdeleted{All 30 targets are pre-selected data points in a large-scale 3D scatter plot from a public dataset. The 3D scatter plot is $10\times10\times8$ m (width $\times$ height $\times$ depth). Participants are instructed to stand at a fixed point five meters away from the scatter plot to maintain consistency in extended arm calibration. Before each study trial, participants take a training session with only 12 targets to practise the required interaction technique and their understanding of Borg CR10 scores. After the training session, participants are told to take a break. They can only start the study trials if they feel refreshed from the break (i.e.\ Borg CR10 $\leq 0.5$). During the study, participants are encouraged to finish the task as quickly as possible to the best of their ability.} Participants are told to take a 15s break after every six targets to maintain the same speed of arm movement.

\begin{figure}
    \centering
    \includegraphics[width = \linewidth]{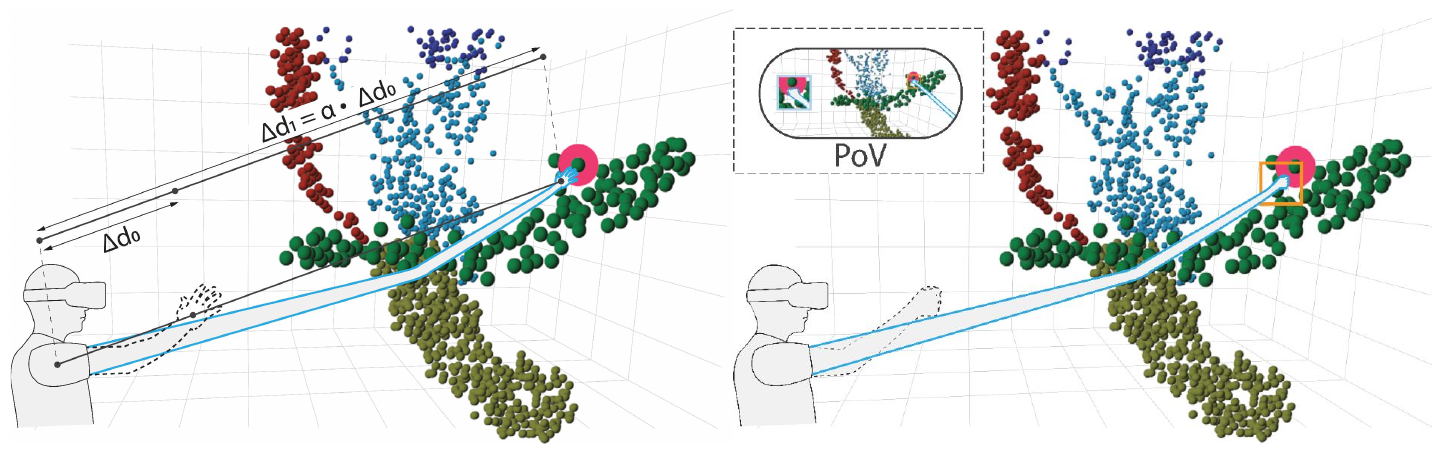}
    \caption{Visualisation of interaction methods used in the external dataset~\cite{magicportal2023}. Left: the extended virtual arm; Right: a pair of portals. }
    \label{fig:s2setup}
\end{figure}

\subsection{Interaction Methods: Extendable Hands and Movable Portals}
\label{sec:AppendixMethods}

The interaction task described above was used with four different interaction methods for distant target selection, described as follows.
\begin{itemize}
    \item Linear Gain: The Linear Gain method amplifies participants’ maximum reach by lengthening the extended virtual arm with a constant scalar ($60 < \alpha < 70$ based on participants’ actual arm length) (see Figure~\ref{fig:s2setup}-left). 
    
    \item Haptic Portal: The setup of the Haptic Portal method includes a non-linear gain, a pair of portals, and a robotic arm-assisted haptic target. The non-linear gain $\alpha$ amplifies participants’ maximum reach with an adaptive scalar based on the real-time hand movement speed $v$ ($cm \cdot s^{-1}$). If $ v < 3$, $\alpha = 5$. Otherwise, $\alpha = 60$. Of the two portals, one is controlled by the extended arm and can be placed anywhere close to the target to enable a clear view of the target location through the other static portal on the left side of the display (see Figure~\ref{fig:s2setup}-right). After successfully selecting the target, participants receive haptic feedback from a robotic arm.

    \item Adaptive Gain (labelled as \textit{High\_Fatigue} in Study 3): A non-linear scalar $\alpha$ applied to the extended virtual arm amplifies participants' maximum reach. This adaptive scalar ($\alpha$) is determined by the hand movement speed. If the physical hand moves slower than a threshold of $3~cm \cdot s^{-1}$,  $\alpha = 5$, otherwise, $\alpha = 60$ (see Figure~\ref{fig:s2setup}-left).

    \item Portal (labelled as \textit{Low\_Fatigue} in Study 3): The test condition featuring a pair of portals ($0.5 m^2 $) that enable a clear view of the target location within the data cluster. One movable portal added to the right hand is controlled by the extended virtual arm. Another static portal serving as the visual reference of the dynamic portal is placed left of the point-of-view (POV) on the display. Participants can first place the movable portal with the extended virtual arm anywhere near the target before reaching out and selecting the target accurately from the portal (see Figure~\ref{fig:s2setup}-right).

\end{itemize}

\end{document}